\documentclass[noshowpacs,amsmath,
twocolumn,
superscriptaddress,
8pt,
aps,prb]{revtex4-2}
\usepackage{setspace}
\usepackage{amsmath}
\usepackage{multirow}
\usepackage{graphicx}

\renewcommand{\figurename}{\textbf{Fig.}}

\usepackage{verbatim}
\usepackage{amsfonts}
\usepackage{amssymb}
\usepackage{epstopdf}
\usepackage{dcolumn}
\usepackage{xcolor}
\usepackage{verbatim}
\usepackage{hyperref}
\usepackage{siunitx}
\usepackage[version=4]{mhchem}
\usepackage{textcomp}
\setcitestyle{super} 
\DeclareGraphicsExtensions{.pdf,.eps,.png,.jpg,.mps}
\begin{document}
\title{Power-efficient ultra-broadband soliton microcombs in resonantly-coupled microresonators}

\author{Kaixuan Zhu$^{1*}$, Xinrui Luo$^{1*}$, Yuanlei Wang$^{1,2*}$, Ze Wang$^{1*}$, Tianyu Xu$^{1}$, Du Qian$^{1}$, Yinke Cheng$^{1,2}$, Junqi Wang$^{1}$, Haoyang Luo$^{1}$, Yanwu Liu$^{1}$, Xing Jin$^{1}$, Zhenyu Xie$^{1}$, Xin Zhou$^{2}$, Min Wang$^{2}$, Jian-Fei Liu$^{2}$, Xuening Cao$^{2}$, Ting Wang$^{2}$, Shui-Jing Tang$^{3}$, Qihuang Gong$^{1,4,5}$, Bei-Bei Li$^{2}$, and Qi-Fan Yang$^{1,4,5\dagger}$\\
$^1$State Key Laboratory for Artificial Microstructure and Mesoscopic Physics and Frontiers Science Center for Nano-optoelectronics, School of Physics, Peking University, Beijing 100871, China\\
$^2$Beijing National Laboratory for Condensed Matter Physics, Institute of Physics, Chinese Academy of Sciences, Beijing 100190, China\\
$^3$National Biomedical Imaging Center, College of Future Technology, Peking University, Beijing, 100871, China\\
$^4$Peking University Yangtze Delta Institute of Optoelectronics, Nantong, Jiangsu 226010, China\\
$^5$Collaborative Innovation Center of Extreme Optics, Shanxi University, Taiyuan 030006, China\\
$^{*}$These authors contributed equally to this work.\\
$^{\dagger}$Corresponding author: leonardoyoung@pku.edu.cn}

\begin{abstract}
The drive to miniaturize optical frequency combs for practical deployment has spotlighted microresonator solitons as a promising chip-scale candidate \cite{Kippenberg2018}. However, these soliton microcombs could be very power-hungry when their span increases, especially with fine comb spacings \cite{yang2024efficient}. As a result, realizing an octave-spanning comb at microwave repetition rates for direct optical-microwave linkage is considered not possible for photonic integration due to the high power requirements. Here, we introduce the concept of resonant-coupling to soliton microcombs to reduce pump consumption significantly. Compared to conventional waveguide-coupled designs, we demonstrate (i) a threefold increase in spectral span for high-power combs and (ii) up to a tenfold reduction in repetition frequency for octave-spanning operation. This configuration is compatible with laser integration and yields reliable, turnkey soliton generation. By eliminating the long-standing pump-power bottleneck, microcombs will soon become readily available for portable optical clocks \cite{newman2019architecture,roslund2024optical,hilton2025demonstration}, massively parallel data links \cite{marin2017microresonator,feldmann2021parallel,shu2022microcomb}, and field-deployable spectrometers \cite{suh2016microresonator,dutt2018chip,yang2019vernier}.
\end{abstract}

\maketitle

\medskip

Two decades after their invention, optical frequency combs are coming out of laboratories to the real world \cite{Lezius:16,roslund2024optical,hilton2025demonstration}. Accelerating this trend demands further reductions in size and power consumption. Soliton microcombs offer a chip-scale solution: generated in high-Q nonlinear microresonators pumped by continuous-wave lasers, they exploit the balance between Kerr nonlinearity and anomalous dispersion to produce repetitive pulse trains, which manifest as phase-coherent teeth equally spaced by the repetition rate in the spectral domain \cite{herr2014temporal,yi2015soliton,brasch2016photonic}. These microcombs hold promise for on-chip optical frequency synthesizers~\cite{spencer2018optical}, clocks~\cite{newman2019architecture}, and spectrometers~\cite{suh2016microresonator,dutt2018chip,yang2019vernier}, and their wide mode spacing (tens of gigahertz) suits wavelength-division multiplexing in communications~\cite{marin2017microresonator,feldmann2021parallel,shu2022microcomb}.

\begin{figure*}[t]
    \centering
    \includegraphics[width=\linewidth]{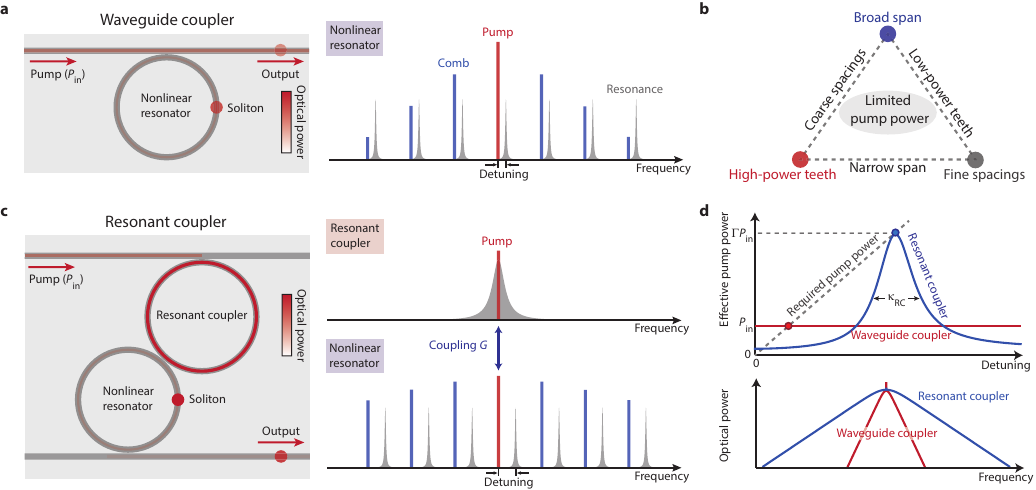} 
    \caption{{\bf Pumping strategies of soliton microcombs.} 
{\bf a,c,} Left: configurations of a nonlinear microresonator pumped via a waveguide coupler ({\bf a}) or a resonant coupler ({\bf c}), with the optical power indicated by color. Right: corresponding diagrams of energy flow. {\bf b,} The ``impossible trinity'' of soliton microcombs under limited pump power. {\bf d,} Top: effective pump power versus detuning. The dashed grey line denotes the minimum pump power required for soliton microcombs. Red and blue dots indicate the maximum detuning for soliton microcombs generated using waveguide couplers and resonant couplers, respectively. Bottom: optical spectra for soliton microcombs at the two detunings, obtained using waveguide couplers (red) and resonant couplers (blue).}
    \label{Fig:Scheme}
\end{figure*}

Key performance metrics of any comb source are its span, power, and spacing. Fine spacing eases direct electrical detection; octave-spanning bandwidth enables carrier-envelope offset measurements through f-2f self-referencing~\cite {telle1999carrier}; high-power teeth maximize data throughput in communications. However, these metrics are usually coupled (Fig.~\ref{Fig:Scheme}a). In conventional soliton microcomb architecture, a nonlinear microresonator (NR) is evanescently coupled to a bus waveguide, in which four-wave mixing initiates when the input power $P_\mathrm{in}$ exceeds the threshold $P_{\mathrm{th}}$, but stable soliton formation further demands red-detuned pump and additional power. The 3-dB bandwidth $\Delta f_{\mathrm{3dB}}$, central-tooth power $P_{\mathrm{c}}$, and repetition rate $f_{\mathrm{r}}$ are constrained by available pump power $P_{\mathrm{in}}$:
\begin{equation}
  \frac{P_{\mathrm{c}}\Delta f_\mathrm{3dB}^2}{f_r^2}\leq 3.1\times\eta_{\mathrm{NR}}^2P_\mathrm{in}.
  \label{Eq:Trinity}
\end{equation}
where $\eta_{\mathrm{NR}} = \kappa_{\mathrm{e,NR}} / \kappa_{\mathrm{NR}}$ is the loading factor of the NR, with $\kappa_{\mathrm{NR}}$ and $\kappa_{\mathrm{e,NR}}$ denoting the NR's dissipation rate and the coupling rate to the waveguide, respectively (see Methods). This ``impossible trinity'' cannot be simultaneously optimized given the limited pump power available from on-chip lasers (Fig.~\ref{Fig:Scheme}b). Also, the quadratic scaling law indicates that increasing the bandwidth or reducing the repetition rate is more challenging than increasing the tooth power.

Several strategies have been proposed to relax this constraint \cite{yang2024efficient}. In particular, resonant couplers (RCs) -- tested in fiber~\cite{xue2019super} and electro-optic resonators~\cite{hu2022high} -- can enhance pump delivery and broaden comb spectra. Here, we demonstrate resonantly-coupled soliton microcombs, achieving up to threefold wider bandwidths and the first octave-spanning soliton combs at microwave repetition rates using only a continuous-wave pump.

\medskip
\noindent{\bf Results}

\noindent{\bf Principle}

\noindent Our architecture, illustrated in Fig.~\ref{Fig:Scheme}c, interposes an auxiliary microresonator (RC) between the bus waveguide and the NR. In this configuration, the RC provides a resonant enhancement of the pump power; the enhanced pump power is delivered to the pump resonance of the NR via the inter-resonator coupling. When the pump laser is tuned to the RC resonance, the effective pump power delivered to the NR is enhanced by a factor on the order of
\begin{equation}
  \Gamma = \frac{4G^2}{\kappa_\mathrm{RC}\,\kappa_\mathrm{NR}},
\end{equation}
where $G$ is the coupling rate between the resonators and $\kappa_{\mathrm{RC}}$ is the RC's dissipation rate (see supplementary materials). To suppress unwanted parametric oscillations in the RC, we typically set $\kappa_{\mathrm{RC}}\gg\kappa_{\mathrm{NR}}$. The enhancement can be considerable for $G \gg \kappa_\mathrm{NR},\,\kappa_\mathrm{RC}$, such that this resonant coupler can outperform a direct waveguide coupler over a bandwidth set by $\kappa_{\mathrm{RC}}$ (Fig.~\ref{Fig:Scheme}d). Since the maximum accessible detuning scales with pump power (dashed grey line in Fig.~\ref{Fig:Scheme}d), we can thus access much larger detunings once the RC resonance is red-detuned relative to the NR resonance. This also dramatically increases the soliton span (scales as $\sqrt{\delta\omega}$), which can also be inferred from Eq. \ref{Eq:Trinity} by applying the enhancement to the pump power.

\begin{figure*}[t]
    \centering
    \includegraphics[width=\linewidth]{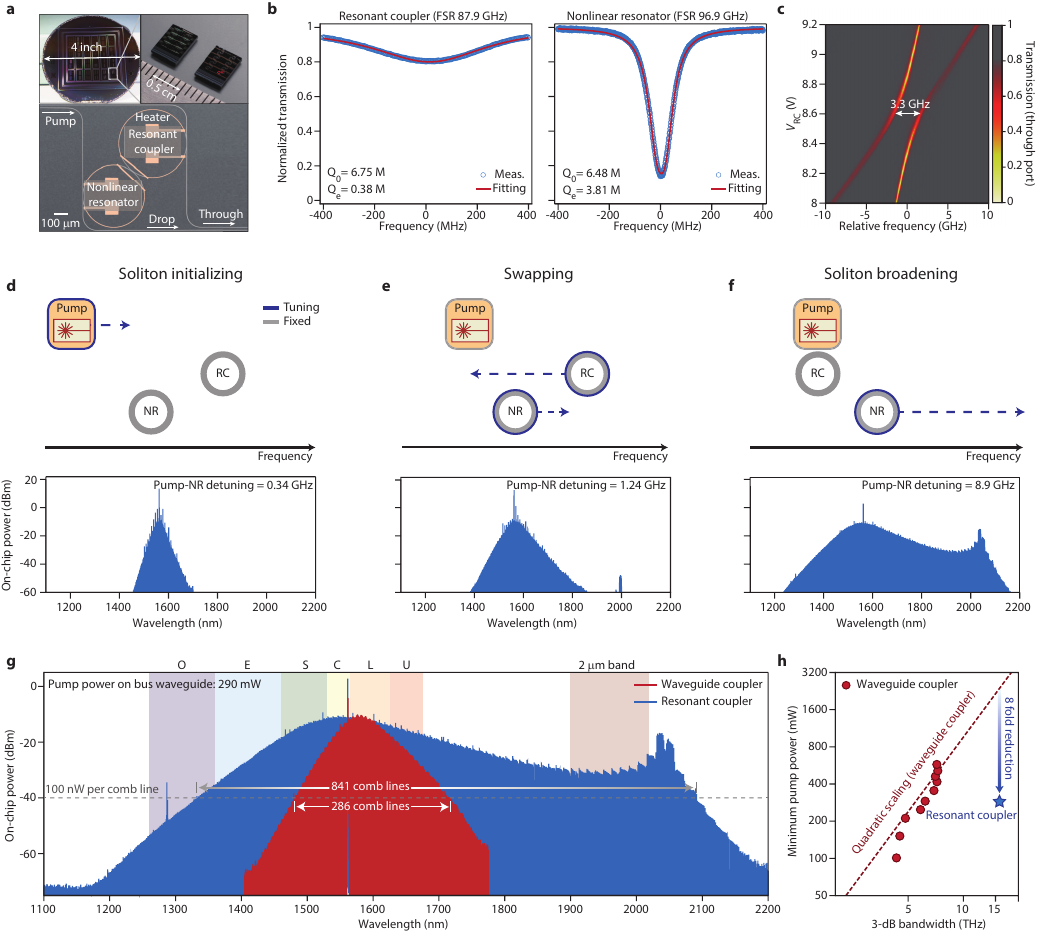}
    \caption{{\bf High-power ultra-broadband soliton microcombs.} 
{\bf a,} Photos of the wafer, chips, and the coupled Si$_3$N$_4$ microresonators. {\bf b,} Measured transmission spectra revealing the intrinsic quality factor $Q_0$ and the external coupling quality factor $Q_\mathrm{e}$ for both the resonant coupler and the nonlinear microresonator. {\bf c,} Transmission spectra from the through port as a function of the voltage ($V_\mathrm{RC}$) applied to the resonant coupler's heater. The minimum frequency difference between the hybridized modes is 3.3 GHz. {\bf d}--{\bf f,} Sequential stages for generating ultra-broadband solitons in a resonantly-coupled NR. Top panel: the relative frequency positions and tuning directions of the pump, RC, and NR. Bottom panel: corresponding optical spectra of soliton microcombs. {\bf g,} Comparison of optical spectra for soliton microcombs generated using conventional waveguide couplers (red) and resonant couplers (blue). All power refers to on-chip power. Communication bands covered by optical amplifiers are highlighted with different color shadings. {\bf h,} Measured minimum pump power as a function of 3-dB bandwidth of soliton microcombs pumped via the waveguide coupler (red dots) and the resonant coupler (blue star) on a log-log scale. The red dashed line represents the quadratic scaling.}
    \label{Fig:Device}
\end{figure*}

\begin{figure*}[t]
\centering
\includegraphics[width=\linewidth]{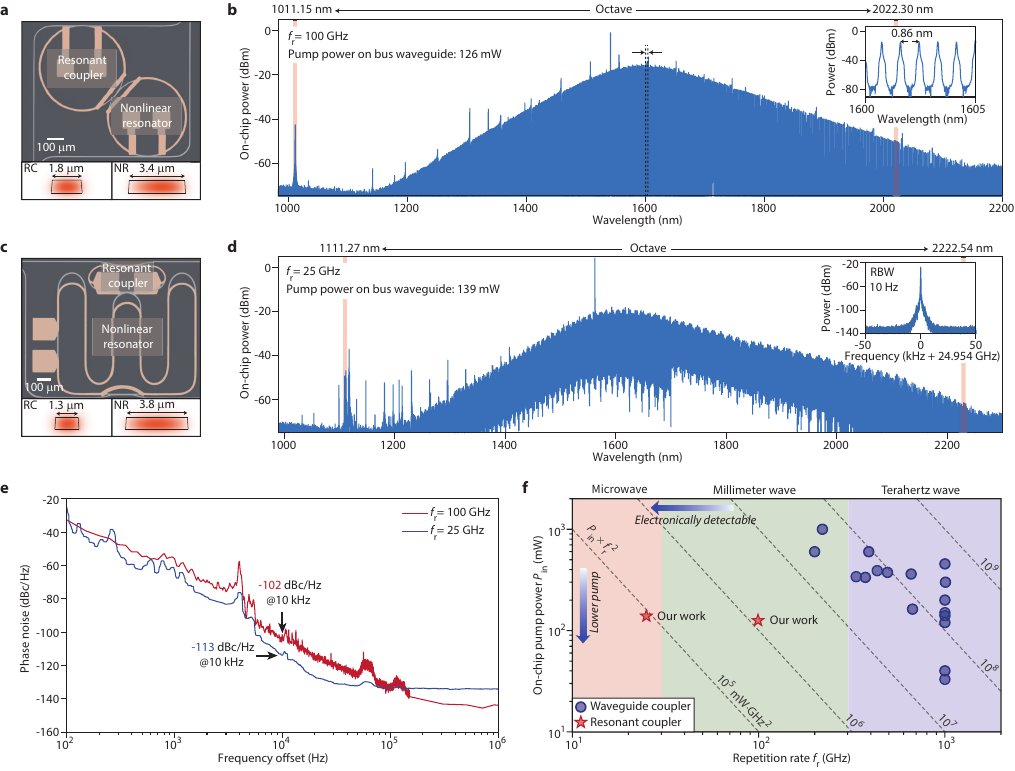}
\caption{{\bf Octave-spanning soliton microcombs at millimeter wave and microwave rates.} 
{\bf a,} Top: image of the coupled ring microresonators. Bottom: the cross-sectional profile of the TE fundamental mode. {\bf b,} Optical spectrum of the octave-spanning soliton microcomb at $f_\mathrm{r}$ of 100 GHz. Insets: zoom-in view of the spectrum between 1600 nm and 1605 nm.
{\bf c,} Top: image of the coupled finger-shaped and racetrack microresonators. Bottom: the cross-sectional profile of the TE fundamental mode. {\bf d,} Optical spectrum of the octave-spanning soliton microcomb at $f_\mathrm{r}$ of 25 GHz. Inset: the electrical beat note at 24.954 GHz. RBW: resolution bandwidth. {\bf e,} Repetition-rate phase noise of 100 GHz (red) and 25 GHz (blue) soliton microcombs. {\bf f,} Comparison of the on-chip pump powers and repetition rates of reported octave-spanning soliton microcombs pumped by continuous-wave lasers. Data from waveguide-coupled configurations are compiled from refs.~\cite{li2017stably,pfeiffer2017octave,weng2025thermally,zang2024foundry,briles2020generating,briles2018interlocking,briles2021hybrid,weng2022dual, liu2021aluminum,weng2021directly, gu2023octave, he2021octave,song2024octave,wang2024octave}.}
\label{Fig:Octave}
\end{figure*}

\begin{figure*}
\centering
\includegraphics{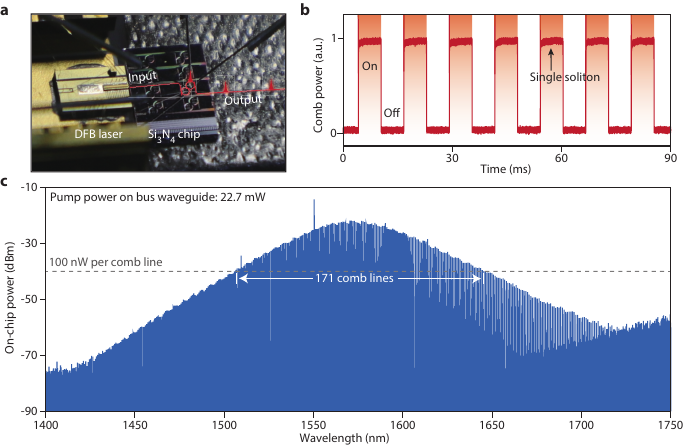}
\caption{{\bf Hybrid-integrated turnkey soliton microcombs.} 
\textbf{a,} Photo showing a DFB laser directly coupled to a \ce{Si3N4} chip. \textbf{b,} Measured comb power when the soliton is turned on and off 7 consecutive times. \textbf{c,} Optical spectrum of a soliton microcomb.}
\label{Fig:Turnkey}
\end{figure*}

\medskip
\noindent{\bf High-power ultra-broadband soliton microcombs}

\noindent We implement our design in 786-nm-thick Si$_3$N$_4$ microresonators fabricated via subtractive processing (see Methods). The RC (waveguide width 1.5 \textmu m) and NR (1.8 \textmu m) have free-spectral ranges (FSRs) of 87.9 GHz and 96.9 GHz, respectively (Fig.~\ref{Fig:Device}a). Both resonators exhibit intrinsic $Q_0\sim7\times10^6$; the RC is overcoupled ($Q_{\mathrm{e}}\approx0.38\times10^6$), while the NR has $Q_{\mathrm{e}}\approx3.8\times10^6$ (Fig.~\ref{Fig:Device}b). Tuning of the resonances is provided by integrated heaters. Adjusting the heater power allows for observation of avoided crossings between two resonances, which reveals inter-resonator coupling $G/2\pi=1.65$ GHz (Fig.~\ref{Fig:Device}c) and predicts $\Gamma\sim100$.

Soliton initiation via the RC differs from conventional schemes (see Methods and supplementary materials). With the RC initially blue-detuned, we sweep the pump into the NR resonance to access single solitons in $\sim80\%$ of trials (Extended Data Fig.~\ref{EXT:Tuning}). At 0.34 GHz detuning, the comb spans 234 nm at -60 dBm (Fig.~\ref{Fig:Device}d). Tuning the RC red and NR blue leads to the swapping of their frequencies, which also increases detuning to 1.24 GHz and broadens the comb span to 488 nm (Fig.~\ref{Fig:Device}e). At the final stage, we red-shift the NR to ~9 GHz detuning to extend the comb span to 937 nm (Fig.~\ref{Fig:Device}f). Note that the spectrum becomes smooth at large detuning as avoided crossings become negligible. Further increasing the detuning would cause modulational instability in the RC, which would destabilize the soliton (see supplementary materials). This reduces the maximum detuning from the theoretical prediction, and could be overcome by further reducing the $Q$ of the RC.

\begin{figure*}[t]
    \centering
    \includegraphics{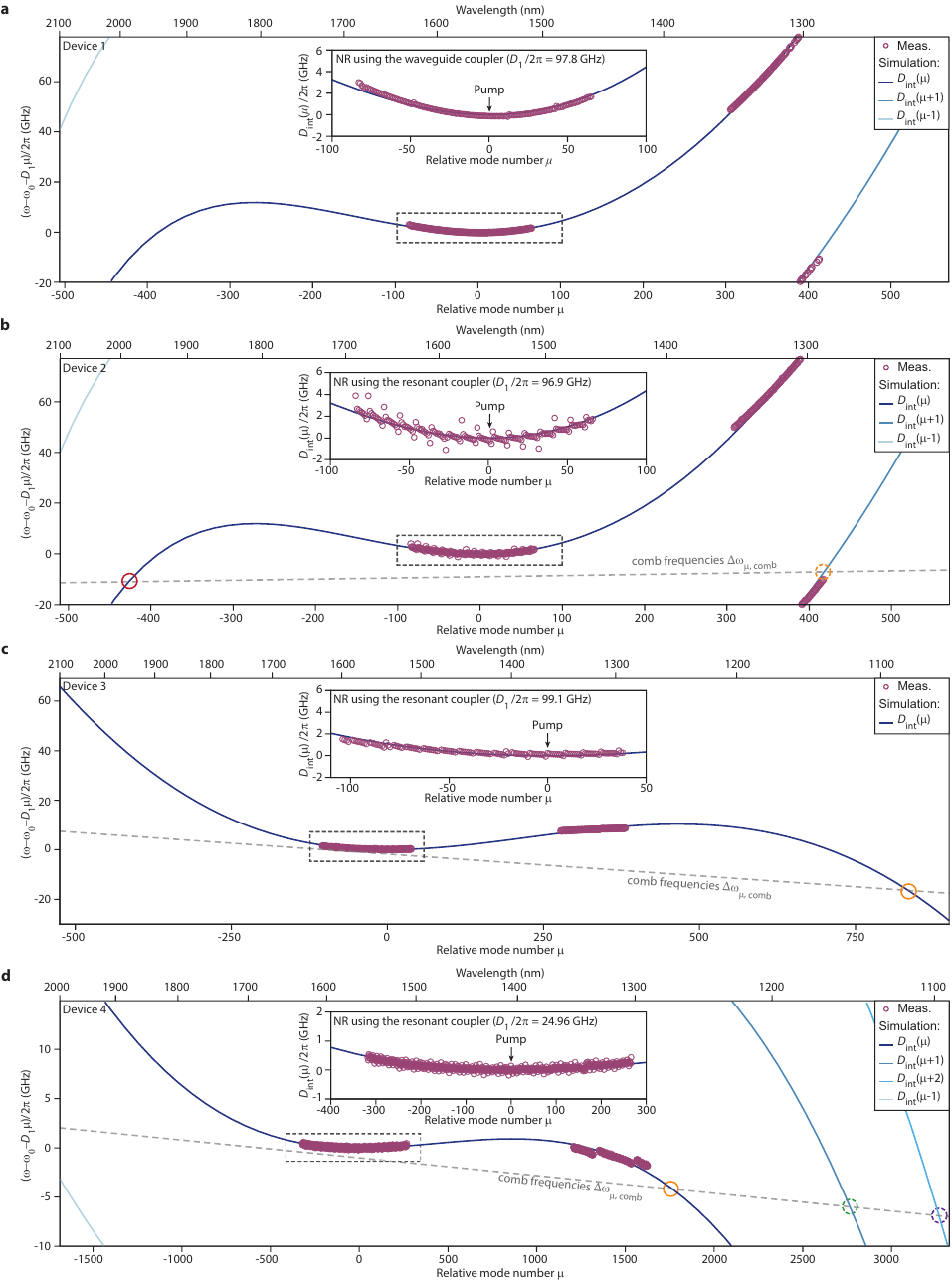}
    \caption{{\bf Mode family dispersion.} Integrated dispersion of the NRs in Device 1-4, respectively. Measured and simulated $D_\mathrm{int}$ values are denoted by purple circles and blue curves, respectively. $D_{\mathrm{int}}(\mu)$ and its shifted values $D_{\mathrm{int}}(\mu\pm n)=D_{\mathrm{int}}(\mu)\mp
    nD_1~(n=1, 2)$ are represented in distinct shades of blue. Dashed grey lines correspond to the hypothetical soliton comb frequencies in the relative frequency frame, with their non-zero slope indicating a repetition rate offset from the NR's FSR. Circles mark dispersive wave locations: solid for phase matching $D_{\mathrm{int}}(\mu)=\Delta\omega_{\mu, \mathrm{comb}}$ and dashed for quasi-phase matching $D_{\mathrm{int}}(\mu+ n)=\Delta\omega_{\mu, \mathrm{comb}}~(n=1, 2)$. Insets: zoom-in view of $D_\mathrm{int}(\mu)$ values near the pump frequency. }
    \label{Dispersion}
\end{figure*}

Owing to the NR’s large group-velocity dispersion (Extended Data Fig.~\ref{Dispersion}b) and strong coupling to the bus waveguide, the comb attains -10.9 dBm tooth power at the center and covers O-band to 2 \textmu m (Fig.~\ref{Fig:Device}g). We record $15.4\,$mW of total drop-port power, corresponding to 5.3\,\% conversion efficiency, with an additional 54\,mW of comb power emitted from the through port due to RC-NR leakage. Benchmarking against a waveguide-coupled NR (identical geometry) shows that at 290 mW pump, the conventional device achieves 6.2-THz 3-dB bandwidth with 286 lines over 100 nW, whereas the device using RC achieves 15.8-THz 3-dB bandwidth with 841 lines over 100 nW (Fig.~\ref{Fig:Device}h). Even with $600\,$mW launched into the bus, the waveguide-coupled device caps at $7.2\,$THz; extrapolating the quadratic pump-span scaling suggests more than 2 W would be needed to match the RC, underscoring 8-fold pump power enhancement afforded by resonant coupling (Fig.~\ref{Fig:Device}h).

\medskip
\noindent{\bf Octave-spanning soliton microcombs at microwave and millimeter-wave rates}

\noindent By widening the NR waveguide (Fig.~\ref{Fig:Octave}a, c), we reduce its group-velocity dispersion that facilitates a broader comb span (Extended Data Fig.~\ref{Dispersion}c, d). In a 100 GHz FSR device (3.4 \textmu m width), 126 mW pump at 1541 nm delivers an octave spectrum from 1007 to 2130 nm (Fig.~\ref{Fig:Octave}b), with coherent dispersive waves near 1011 nm confirmed by heterodyning against another laser (see supplementary materials). In a 25 GHz device (finger-shaped NR and racetrack RC, 1 mm$^2$ footprint), 139 mW pump at 1562 nm generates an octave spectrum from 1098 to 2250 nm (Fig.~\ref{Fig:Octave}d). Direct photodetection of the soliton microcomb produces a monotone electrical beatnote that corresponds to the repetition rate. Both spectra exhibit notable shifts of the spectral-envelope center from the pump wavelength, which is driven by Raman self-frequency shifts \cite{karpov2016raman,yi2016theory,wang2018stimulated} and dispersive-wave recoil \cite{brasch2016photonic,yi2017single}. This limits the maximum comb span and must be balanced to realize even broader combs \cite{wang2018stimulated}.

\begin{figure*}[t]
\centering
\includegraphics{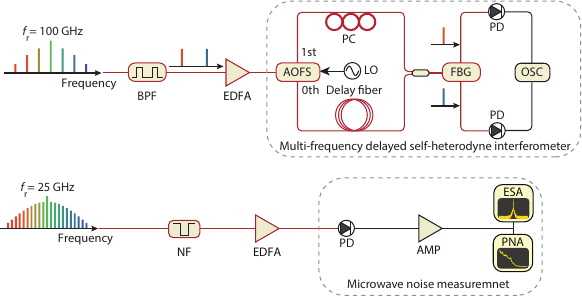}
\caption{{\bf Experimental setup for coherence characterization of octave-spanning soliton microcombs.}
BPF: band-pass filter; EDFA: erbium-doped fiber amplifier; AOFS: acousto-optic frequency shifters. PD: photodetector;  LO: local oscillator; FBG: fiber Bragg grating; OSC: oscilloscope; NF: notch filter; AMP: electrical amplifier; ESA: electrical spectral analyzer; PNA: phase noise analyzer. }
\label{EXT:Cohence}
\end{figure*}

To quantify coherence for future self-referencing, we measure the phase noise of the 100 GHz comb using a multi-frequency delayed self-heterodyne interferometer, and that of the 25 GHz comb with a commercial phase-noise analyzer (Extended Data Fig.~\ref{EXT:Cohence}; see Methods). At a 10 kHz offset, we record -102 dBc/Hz (100 GHz) and –113 dBc/Hz (25 GHz), comparable to the lowest reported for free-running integrated soliton microcombs (Fig.~\ref{Fig:Octave}e).

Figure~\ref{Fig:Octave}f plots the on-chip pump power required for octave-spanning comb generation versus repetition rate for various CW-pumped platforms. Because $P_\mathrm{in}\propto f_\mathrm{r}^{-2}$ (Eq.~\ref{Eq:Trinity}), we compare the figure of merit \(P_\mathrm{in}\times f_\mathrm{r}^2\). Our RC architecture achieves values around \(10^{5}\)\,mW\(\cdot\)GHz$^2$, which is lower than the best results reported in conventional waveguide-coupled couplers by two orders of magnitude.

\medskip
\noindent{\bf Hybrid-integrated turnkey soliton microcombs}

\noindent We use an on-chip laser to drive the soliton microcomb through an RC. A distributed-feedback (DFB) laser is coupled into the Si$_3$N$_4$ chip, delivering approximately 20 mW of optical power to the bus waveguide (Fig.~\ref{Fig:Turnkey}a). To enable operation at this low pump power, an NR with a higher $Q$ is selected (see Methods). Without an optical isolator, light backscattered from the microresonator re-enters the laser cavity and perturbs its tuning. This phenomenon, known as self-injection locking\cite{liang2015high,raja2019electrically,shen2020integrated,jin2021hertz}, narrows the laser’s linewidth and biases the system toward soliton microcomb generation when the feedback phase is appropriately tuned. In our implementation, the reinjection feedback phase is adjusted using a piezoelectric stage to meet the condition for stable soliton formation. 

By optimizing the feedback phase, single-soliton microcombs emerge deterministically each time the laser current is tuned to a predetermined setpoint. To emulate soliton turn-on dynamics, the laser current is modulated with a square wave (Fig.~\ref{Fig:Turnkey}b). Each time the current is switched to the target value, a single-soliton state reliably forms in the NR. Ultimately, this self-injection-locked pumping approach obtains single-soliton microcombs with a $f_\mathrm{r}$ of 99 GHz and an optical bandwidth exceeding 300 nm (Fig.~\ref{Fig:Turnkey}c). 171 comb lines are above 100 nW. This is the broadest soliton microcombs at such a repetition rate when pumped by on-chip lasers.

\medskip
\noindent{{\bf Discussion and outlook}}

\noindent Generating octave-spanning microcombs at electronically accessible rates with low pump power is a milestone and can unlock many opportunities. First, optical frequency division \cite{xie2017photonic}, optical frequency synthesis, and optical clocking based on self-referenced optical frequency combs can be implemented on chip without complex protocols involving multiple combs \cite{spencer2018optical,newman2019architecture}. Second, these combs can serve as precise wavelength calibrators in astronomical spectrographs, delivering densely spaced, ultra-stable lines across the visible and near-infrared \cite{suh2019searching,obrzud2019microphotonic}. Third, in the time domain they produce sub-20-fs pulses, which can be used for synthesizing low-duty-cycle femtosecond pulse trains and arbitrary optical waveforms. These applications were previously difficult to operate or fell short in performance with narrow-band microcombs.

With the pump-power bottleneck relaxed, future efforts can concentrate on optimizing microcomb performance through engineering the NRs. Large-dispersion NRs within the resonant-coupler architecture will deliver large numbers of high-power comb teeth with signal-to-noise ratios suited to advanced telecom formats -- potentially obviating external optical amplification \cite{jorgensen2022petabit,zhang2023clone}. Simultaneously, robust f-2f self-referencing will demand precise control over dispersive-wave generation \cite{li2017stably,pfeiffer2017octave,drake2019terahertz,liu2021aluminum,song2024octave}. Finally, just as the input port of the microcomb can be engineered to enhance pump delivery, the output spectrum itself may be sculpted via wavelength-selective couplers to meet specific application requirements.

\bigskip
\noindent{{\bf Methods}}

\begin{footnotesize}
\noindent{\bf Impossible trinity of soliton microcombs.}
The conventional soliton microcombs are described by the Lugiato–Lefever equation~\cite{lugiato1987spatial}:
\begin{equation}
       \frac{\partial A}{\partial T}  = -\frac{\kappa_{\mathrm{NR}}}{2}A - i\delta\omega A + i\frac{D_{\mathrm{2}}}{2}\frac{\partial^2 A}{\partial \phi^2} +ig|A|^2A + \sqrt{\frac{\kappa_\mathrm{e,NR} P_\mathrm{in}}{\hbar\omega_0}},
\end{equation}
where $T$ is the slow time (lab time) and $\phi$ is the angular coordinate in the moving frame. $A(T, \phi)$ corresponds to the slowly varying field amplitude, which is normalized such that $|A|^2$ corresponds to the intracavity photon number. $D_{\mathrm{2}}$ is the second-order dispersion. The decay rates of NR is defined as $\kappa_{\mathrm{NR}} = \kappa_{\mathrm{0,NR}} + \kappa_{\mathrm{e,NR}}$, where $\kappa_{\mathrm{0,NR}}$ is the intrinsic decay rates and $\kappa_{\mathrm{e,NR}}$ is the coupling rates to the waveguide. $g$ denotes the nonlinear coefficient, which is defined as $g=\frac{\hbar \omega_0^2 cn_2}{n_0^2 V_{\mathrm{eff}}}$, where $V_{\mathrm{eff}}$ is the effective mode volume and $n_2$ is the nonlinear refractive index associated with the refractive index $n_0$. $\delta\omega$ is the pump-NR detuning and $P_\mathrm{in}$ is the input pump power. The onset of four-wave mixing occurs when $P_\mathrm{in}$ exceeds the threshold $P_{\mathrm{th}}$:
\begin{equation}
    P_{\mathrm{th}}= \frac{\hbar \omega_0 \kappa_{\mathrm{NR}}^3}{8g\kappa_{\mathrm{e, NR}}}
\end{equation}
However, sustaining solitons at a given detuning $\delta\omega$ requires additional pump power, 
\begin{equation}
    P_\mathrm{in}\geq \frac{16}{\pi^2}\times \frac{\delta\omega P_\mathrm{th}}{\kappa_\mathrm{NR}}, 
    \label{eq:Pin-detuning}
\end{equation}
The detuning is a key parameter determining the comb span:
\begin{equation}
    \Delta f_{\mathrm{3dB}}=\frac{1.763}{\pi^2}\times\sqrt{-\frac{2n_0\delta\omega}{c\beta_{\mathrm{2}}}}, 
    \label{eq:band-detuning}
\end{equation}
where $\beta_{\mathrm{2}}=-\frac{n_0D_\mathrm{2}}{cD^2_\mathrm{1}}$ is the group velocity dispersion coefficient, with $D_\mathrm{1}$ denoting the FSR in angular frequency. Combining Eqs.~\ref{eq:Pin-detuning},~\ref{eq:band-detuning} and using the approximation $D_{\mathrm{1}}\approx2\pi f_{\mathrm{r}}$, we derive a lower bound on the pump power required to support a soliton microcomb with a specified bandwidth:
\begin{equation}
    P_{\mathrm{in}} \geq -\frac{\pi^2}{1.763^2}\times\frac{\kappa_{\mathrm{NR}}\beta_{\mathrm{2}}}{\eta_{\mathrm{NR}}\gamma}\times\frac{\Delta f_{\mathrm{3dB}}^2}{f_\mathrm{r}}, 
    \label{eq:Pmin}
\end{equation}
where $\gamma=\frac{\omega_0n_2}{cA_\mathrm{eff}}$ is the nonlinear parameter, with $A_\mathrm{eff}$ denoting the effective mode area. Additionally, the central-tooth power ($P_\mathrm{c}$) of soliton microcombs can be expressed as: 
\begin{equation}
    P_\mathrm{c} =-\pi^2\times\frac{\kappa_{\mathrm{e, NR}}\beta_{\mathrm{2}} f_{\mathrm{r}}}{\gamma}.
    \label{eq:Pcen}
\end{equation}
By eliminating the material-dependent terms ($\beta_{\mathrm{2}}, \gamma$) through dividing Eq.~\ref{eq:Pmin} by Eq.~\ref{eq:Pcen}, we obtain a constraint among central-tooth power, 3-dB bandwidth and repetition rate under limited pump power-- referred to as the "impossible trinity", 
\begin{equation}
     \frac{P_{\mathrm{c}}\Delta f_{\mathrm{3dB}}^2}{f_{\mathrm{r}}^2} \leq 1.763^2\times \eta_\mathrm{NR}^2P_ {\mathrm{in}} \approx3.1\times \eta_\mathrm{NR}^2P_ {\mathrm{in}} 
\end{equation}
A detailed derivation is provided in the supplementary materials.

\medskip
\noindent{\bf Device fabrication.}
The Si$_3$N$_4$ coupled microresonators are fabricated on a 4-inch wafer through subtractive processes \cite{wang2025compact}. Initially, a 786~nm-thick Si$_3$N$_4$ film is deposited in two steps onto a wet-oxidized silicon substrate featuring stress-release patterns. Electron beam lithography is used to define the pattern, followed by dry etching to transfer the resist pattern to the Si$_3$N$_4$ film. The wafer is then annealed at 1200$^{\circ}$C to remove residual N-H and Si-H bonds from the Si$_3$N$_4$ film. SiO$_2$ cladding is deposited, followed by a second annealing step for densifying the film. A lift-off process is then used to define the heater patterns. Finally, the wafer is diced into 5 mm $\times$ 5 mm chips. 

\medskip
\noindent{\bf Device characterization.}
Our work involves five devices for microcomb generation. Device 1 is used for soliton generation in the waveguide-coupled NR (Fig.~\ref{Fig:Device}h). Device 2 is employed to demonstrate high-power ultra-broadband soliton microcombs (Fig.~\ref{Fig:Device}g). Device 3 supports an octave-spanning microcomb at $f_\mathrm{r}$ of 100 GHz (Fig.~\ref{Fig:Octave}a). Device 4 enables an octave-spanning microcomb at $f_\mathrm{r}$ of 25 GHz (Fig.~\ref{Fig:Octave}c). Device 5 realizes hybrid-integrated soliton microcombs (Fig.~\ref{Fig:Turnkey}a). Devices 2-5 adopt the RC architecture.

For Device 1, the NR with waveguide width 1.8 \textmu m has $Q_{0}=7.29\times 10^6$, $Q_{\mathrm{e}}=3.83\times 10^6$. For Device 2, the NR (1.8 \textmu m) has $Q_{0}=6.48\times 10^6$, $Q_{\mathrm{e}}=3.81\times 10^6$, while the RC (1.5 \textmu m) has $Q_{0}=6.75\times 10^6$, $Q_{\mathrm{e}}=0.38\times 10^6$. The inter-resonator coupling rate is $ G/2\pi = 3.3$ GHz. For Device 3, the NR (3.4 \textmu m) has $Q_{0}=20.64\times 10^6$, $Q_{\mathrm{e}}=3.29\times 10^6$, while the RC (1.8 \textmu m) has $Q_{0}=6.67\times 10^6$, $Q_{\mathrm{e}}=0.25\times 10^6$. The inter-resonator coupling rate is $ G/2\pi = 0.73$ GHz. For Device 4, the NR (3.8 \textmu m) has $Q_{0}=9.56\times 10^6$, $Q_{\mathrm{e}}=9.56\times 10^6$, while the RC (1.3 \textmu m) is strongly overcoupled, making its Q factor difficult to characterize. The inter-resonator coupling rate is $G/2\pi =0.32$ GHz. For Device 5, the NR (2.5 \textmu m) has $Q_{0}=14.52\times 10^6$, $Q_{\mathrm{e}}=21.44\times 10^6$, while the RC (1 \textmu m) has $Q_{0}=2.79\times 10^6$, $Q_{\mathrm{e}}=0.2\times 10^6$. The inter-resonator coupling rate is $ G/2\pi = 0.47$ GHz.

To characterize the dispersive wave position, we measure the broadband dispersion of NRs by sweeping several widely tunable lasers (Toptica CTL series) across the resonances while recording the transmission signal with a photodetector. For frequency calibration, part of the laser power is split before entering the microresonator and routed through an unbalanced Mach–Zehnder interferometer (UMZI), which generates a sinusoidal reference signal. The FSR and dispersion of the UMZI in each spectral band are calibrated using a vector spectrum analyzer \cite{Luo2024wideband}. We also perform finite element simulations of broadband dispersion for each device using its respective geometry. Both the simulated dispersion and experimental data are shown in Extended Data Fig.~\ref{Dispersion}. In the plots, the integrated dispersion is defined as $D_\mathrm{int}(\mu)=\omega_{\mu}-\omega_0-\mu D_1$, where $\omega_{\mu} $ is the resonant frequency of the $\mu_\mathrm{th}$ mode, and $D_1$ is the FSR in angular frequency. The integrated dispersion of optical modes shifted by $nD_1$ is given by $D_{\mathrm{int}}(\mu\pm n)=D_{\mathrm{int}}(\mu)\mp nD_1$, where $n$ is an integer. The hypothetical soliton comb frequencies in the relative frequency frame are given by $\Delta\omega_{\mu,\mathrm{comb}}=\mu\omega_{\mathrm{r}}+\omega_\mathrm{p}-\omega_0-D_1\mu$, where $\omega_0-\omega_\mathrm{p}$ represents the pump-NR detuning and $\omega_\mathrm{r}$ is the soliton repetition rate~\cite{yi2017single}. The phase-matched and quasi-phase-matched locations of dispersive waves are predicted using the conditions $D_{\mathrm{int}}(\mu)=\Delta\omega_{\mu, \mathrm{comb}}$ and $D_{\mathrm{int}}(\mu+ n)=\Delta\omega_{\mu, \mathrm{comb}}~(n=1, 2)$, respectively~\cite{brasch2016photonic,anderson2022zero}. The predicted dispersive wave location does not match perfectly with the experiment, which can be attributed to the residual difference between the simulated and actual dispersion profiles.

\medskip
\noindent{\bf Characterization of tuning process.}
We employ a probe laser to visualize the tuning dynamics (Extended Data Fig.~\ref{EXT:Tuning}a). Coupled from either the through port or the drop port and collected at the input, the probe signal interrogates the hybridized resonances of the NR and RC. After soliton initiation, we scan the frequency of the probe laser and observe a narrow beatnote resulting from interference with backreflected pump (Extended Data Fig.~\ref{EXT:Tuning}c-e). Broader peaks correspond to hybridized resonances, from which the RC and NR resonant frequencies are deduced (see Supplementary Materials).

\medskip
\noindent{\bf Characterization of repetition-rate noise.}
The repetition-rate phase noise of the 100 GHz soliton microcomb is characterized using a multi-frequency delayed self-heterodyne setup~\cite{lao2023quantum} (Extended Data Fig.~\ref{EXT:Cohence}). Two comb lines are selected by a programmable optical filter and amplified by an EDFA. One path is frequency-shifted, while the other is temporally delayed. After recombination, the signals include both the original and frequency-shifted components of the selected comb lines. These are separated using a fiber Bragg grating and individually detected by two photodetectors. The beatnote phases $\Phi_i, \Phi_j$ for comb modes $i$ and $j$ are simultaneously extracted via the Hilbert transform of the oscilloscope traces. The repetition-rate phase noise is derived from the power spectral density (PSD) of the phase difference between the two beatnotes,
\begin{equation}
    S_{\varphi}(f) = \frac{PSD\left[\Phi_i - \Phi_j\right]}{\left(i-j\right)^2}\frac{1}{4\sin^2\pi f \tau_{\mathrm{d}}},
\end{equation}
where $\tau_{\mathrm{d}}$ denotes the time delay between the two interferometer arms. To prevent singularities at offset frequencies where $\sin{\pi f \tau_{\mathrm{d}}}=0$, a cut-off frequency is set at $1/\tau_{\mathrm{d}}$ and only data points at offset frequencies of $(N+1/2)/\tau_{\mathrm{d}}$, where $N$ is an integer, are retained for analysis.

The repetition rate of the 25 GHz soliton microcomb is directly measured using a high-speed photodetector connected to an electrical spectrum analyzer (Extended Data Fig.~\ref{EXT:Cohence}). Before detection, residual pump light is suppressed using a notch filter, and the comb is amplified to 2 mW. The phase noise is further characterized using a phase noise analyzer (PNA; Rohde \& Schwarz FSWP50).

\end{footnotesize}

\bibliography{ref}

\medskip
\noindent\textbf{Acknowledgments}

\begin{footnotesize}
\noindent This work was supported by National Key R\&D Plan of China (Grant No. 2023YFB2806702), Beijing Natural Science Foundation (Z210004), and National Natural Science Foundation of China (92150108). The authors thank Jincheng Li, Zhigang Hu, Hao Yang, Ruokai Zheng, and Xiaoxuan Peng for assistance in fabrication. The fabrication in this work was supported by the Peking University Nano-Optoelectronic Fabrication Center, Micro/nano Fabrication Laboratory of Synergetic Extreme Condition User Facility (SECUF), Songshan Lake Materials Laboratory, and the Advanced Photonics Integrated Center of Peking University. 
\end{footnotesize}
\medskip

\noindent\textbf{Author contributions} 

\begin{footnotesize}
\noindent Experiments were conceived and designed by K.Z., X.L., and Q.-F.Y. Measurements and data analysis were performed by K.Z., X.L., with assistance from Z.W. and T.X. Numerical simulations were performed by K.Z. and X.L. The device was designed by K.Z. and X.L. with assistance from Y.L. The device was fabricated by Y.W., with assistance from Y.C., J.W., H.L., X.Z., M.W., J.-F.L., X.C., T.W., and B.-B.L. The project was supervised by Q.G. and Q.-F.Y. All authors participated in preparing the manuscript.
\end{footnotesize}

\medskip

\noindent\textbf{Competing interests}

\begin{footnotesize}
\noindent The authors declare no competing interests.
\end{footnotesize}

\medskip

\noindent\textbf{Data availability}

\begin{footnotesize}
\noindent The data that support the plot within this paper and other findings of this study are available upon publication. 
\end{footnotesize}

\medskip

\noindent\textbf{Code availability}

\begin{footnotesize}
\noindent The codes that support the findings of this study are available upon publication. 
\end{footnotesize}

\medskip

\noindent\textbf{Additional information}

\begin{footnotesize}
\noindent Correspondence and requests for materials should be addressed to Q-F.Y.
\end{footnotesize}

\renewcommand{\figurename}{\textbf{Extended Data Fig.}}

\begin{figure*}
\setcounter{figure}{0}
\centering
\includegraphics{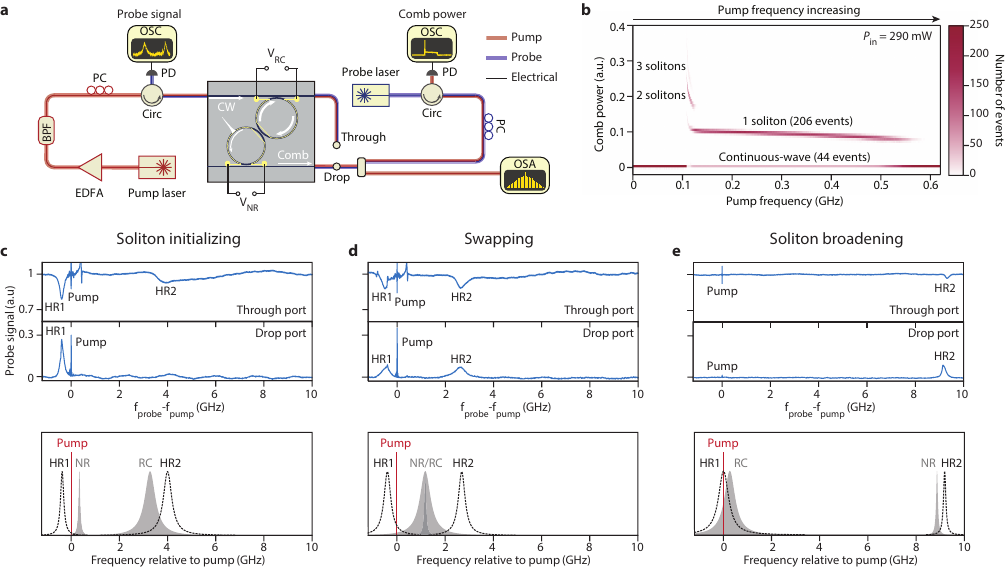}
\caption{{\bf Tuning process.} 
{\bf a,} Experimental setup. EDFA: erbium-doped fiber amplifier; BPF: band-pass filter; PC: polarization controller; Circ: circulator; PD: photodetector; OSC: oscilloscope; OSA: optical spectrum analyzer.
{\bf b,} Comb power versus pump laser frequency for 250 consecutive scans. The number of events is indicated by color.
{\bf c}--{\bf e,} Sequential stages for generating ultra-broadband solitons in a resonantly-coupled NR. Top panel: recorded probe signals when the probe laser is launched from the through and drop ports. The hybridized resonances (HRs) and the beat note with the backreflected pump are indicated. Bottom panel: reconstructed frequencies of NR and RC relative to the pump.}
\label{EXT:Tuning}
\end{figure*}

\end{document}


\title{Supplementary information: Power-efficient ultra-broadband soliton microcombs in resonantly-coupled microresonators}

\author{Kaixuan Zhu$^{1*}$, Xinrui Luo$^{1*}$, Yuanlei Wang$^{1,2*}$, Ze Wang$^{1*}$, Tianyu Xu$^{1}$, Du Qian$^{1}$, Yinke Cheng$^{1,2}$, Junqi Wang$^{1}$, Haoyang Luo$^{1}$, Yanwu Liu$^{1}$, Xing Jin$^{1}$, Zhenyu Xie$^{1}$, Xin Zhou$^{2}$, Min Wang$^{2}$, Jian-Fei Liu$^{2}$, Xuening Cao$^{2}$, Ting Wang$^{2}$, Shui-Jing Tang$^{3}$, Qihuang Gong$^{1,4,5}$, Bei-Bei Li$^{2}$, and Qi-Fan Yang$^{1,4,5\dagger}$\\
$^1$State Key Laboratory for Artificial Microstructure and Mesoscopic Physics and Frontiers Science Center for Nano-optoelectronics, School of Physics, Peking University, Beijing 100871, China\\
$^2$Beijing National Laboratory for Condensed Matter Physics, Institute of Physics, Chinese Academy of Sciences, Beijing 100190, China\\
$^3$National Biomedical Imaging Center, College of Future Technology, Peking University, Beijing, 100871, China\\
$^4$Peking University Yangtze Delta Institute of Optoelectronics, Nantong, Jiangsu 226010, China\\
$^5$Collaborative Innovation Center of Extreme Optics, Shanxi University, Taiyuan 030006, China\\
$^{*}$These authors contributed equally to this work.\\
$^{\dagger}$Corresponding author: leonardoyoung@pku.edu.cn}

\date{\today}


\maketitle

\tableofcontents

\newpage

\section{Theory for waveguide-coupled microresonators}
\label{sec:singletheory}
We begin by analyzing a conventional configuration in which a nonlinear microresonator (NR) is evanescently coupled to a single bus waveguide. When the NR is pumped by with a continuous-wave (CW) laser, the dynamics are governed by the Lugiato–Lefever equation (LLE)~\cite{lugiato1987spatial}:
\begin{equation}
       \frac{\partial A}{\partial T}  = -\frac{\kappa_{\mathrm{NR}}}{2}A - i\delta\omega_{\mathrm{NR}} A + i\frac{D_{\mathrm{2,NR}}}{2}\frac{\partial^2 A}{\partial \phi^2} +ig_{\mathrm{NR}}|A|^2A + \sqrt{\frac{\kappa_\mathrm{e,NR} P_\mathrm{in}}{\hbar\omega_0}},
       \label{eq:lle_NR0}
\end{equation}
where $T$ is the slow time (lab time) and $\phi$ is the angular coordinate in the moving frame. $A(T, \phi)$ corresponds to the slowly varying field amplitude, which is normalized such that $|A|^2$ corresponds to the intracavity photon number. $D_{\mathrm{2, NR}}$ is the second-order dispersion in the NR. The decay rates of NR is defined as $\kappa_{\mathrm{NR}} = \kappa_{\mathrm{0,NR}} + \kappa_{\mathrm{e,NR}}$, where $\kappa_{\mathrm{0,NR}}$ is the intrinsic decay rates and $\kappa_{\mathrm{e,NR}}$ is the coupling rates to the waveguide. $g_{\mathrm{NR}}$ denotes the nonlinear coefficient of the NR, which is defined as $g_{\mathrm{NR}}=\frac{\hbar \omega_0^2 cn_2}{n_0^2 V_{\mathrm{eff, NR}}}$, where $V_{\mathrm{eff, NR}}$ is the effective mode volume of NR and $n_2$ is the nonlinear refractive index associated with the refractive index $n_0$. $\delta\omega_{\mathrm{NR}}$ is the pump-NR detuning and $P_\mathrm{in}$ is the pump power.

The approximate solution for the single-soliton state is given by~\cite{herr2014temporal}, 
\begin{equation}
    A_{\mathrm{tot}} = A_{\mathrm{cw}}+A_{\mathrm{sol}}=A_{\mathrm{cw}}+\sqrt{\frac{2\delta\omega_{\mathrm{NR}}}{g_{\mathrm{NR}}}}\mathrm{sech}\left(\sqrt{\frac{2\delta\omega_{\mathrm{NR}}}{D_{\mathrm{2,NR}}}}\phi\right)e^{i\varphi_0}.
\end{equation}
Here, $A_{\mathrm{cw}}$ is the continuous-wave background and $A_{\mathrm{sol}}$ denotes the hyperbolic secant waveform in the time domain. $\varphi_0$ specifies the soliton's phase relative to the pump, which satisfies:
\begin{equation}
    \cos(\varphi_0) = \frac{4}{\pi}\sqrt{\frac{\delta\omega_{\mathrm{NR}}P_{\mathrm{th, NR}}}{\kappa_{\mathrm{NR}}P_{\mathrm{in}}}},
\end{equation}
where $P_{\mathrm{th,NR}}$ is the parametric oscillation threshold, defined as $P_{\mathrm{th, NR}}= \frac{\hbar \omega_0 \kappa_{\mathrm{NR}}^3}{8g_{\mathrm{NR}}\kappa_{\mathrm{e, NR}}}$. The constraint $\cos(\varphi_0)\leq 1$ implies the minimum pump power required to support a soliton at a given detuning:
\begin{equation}
    P_{\mathrm{in}}\geq \frac{16}{\pi^2}\times\frac{\delta\omega_{\mathrm{NR}}P_{\mathrm{th,NR}}}{\kappa_{\mathrm{NR}}}.
    \label{eq:Pmin-detuning}
\end{equation}
The spectral envelope of the soliton microcomb is given by the Fourier transform:
\begin{equation}
    \tilde{A}_{\mathrm{sol}}(\mu) = \mathcal{F}[A_{\mathrm{sol}}(\phi)]=\sqrt{\frac{D_{\mathrm{2,NR}}}{4g_{\mathrm{NR}}}}\mathrm{sech}\left( \frac{\pi\mu}{2}\sqrt{\frac{D_{\mathrm{2,NR}}}{2\delta\omega_{\mathrm{NR}}}} \right) e^{i\varphi_0}, 
\end{equation}
where $\mu$ is the mode number relative to the pump. Using the relation for comb frequencies, 
\begin{equation}
    \omega_\mu=\omega_\mathrm{p}+\mu \omega_\mathrm{r},
\end{equation}
where $\omega_\mathrm{p}$ is the pump frequency and $\omega_\mathrm{r}$ is the repetition frequency. Since $\omega_\mathrm{r}$ is close to the free spectral range (FSR, $D_{\mathrm{1, NR}}$), $\tilde{A}_{\mathrm{sol}}(\mu)$ becomes:
\begin{equation}
    \tilde{A}_{\mathrm{sol}}(\omega_\mu-\omega_\mathrm{p})=\sqrt{\frac{D_{\mathrm{2,NR}}}{4g_{\mathrm{NR}}}}\mathrm{sech}\left( \frac{\omega_\mu-\omega_\mathrm{p}}{\Delta\omega} \right) e^{i\varphi_0}~~~ \mathrm{with}~ ~\Delta\omega=\frac{2D_{\mathrm{1, NR}}}{\pi}\sqrt{\frac{2\delta\omega_{\mathrm{NR}}}{D_{\mathrm{2,NR}}}}.
\end{equation}
This equation can be reformulated in terms of the group velocity dispersion coefficient $\beta_{\mathrm{2, NR}} = -\frac{n_0D_\mathrm{2, NR}}{cD^2_\mathrm{1, NR}}$ and the nonlinear parameter $\gamma_{\mathrm{NR}}=\frac{\omega_0n_2}{cA_\mathrm{eff, NR}}$, with $A_\mathrm{eff, NR}$ denoting the effective mode area of NR. Using the approximation $D_{\mathrm{1,NR}}\approx2\pi f_{\mathrm{r}}$, where $f_{\mathrm{r}}$ is the soliton repetition rate, the soliton spectrum takes the form: 
\begin{equation}
    \tilde{A}_{\mathrm{sol}}(\omega_\mu-\omega_\mathrm{p})=\pi\sqrt{-\frac{\beta_{\mathrm{2,NR}}f_{\mathrm{r}}}{\hbar\omega_0\gamma_\mathrm{NR}}}\mathrm{sech}\left( \frac{\omega_\mu-\omega_\mathrm{p}}{\Delta\omega} \right) e^{i\varphi_0}~~~ \mathrm{with}~ ~\Delta\omega=\frac{2}{\pi}\sqrt{-\frac{2n_0\delta\omega_{\mathrm{NR}}}{c\beta_{\mathrm{2,NR}}}}.
    \label{eq:solspec}
\end{equation}
Equation~\ref{eq:solspec} reveals two key spectral metrics of soliton microcombs:

1.~Central-tooth power ($P_{\mathrm{c}}$):
\begin{equation}
    P_{\mathrm{c}}= \hbar\omega_0\kappa_{\mathrm{e,NR}}|\tilde{A}_{\mathrm{sol}}(0)|^2=-\pi^2\times\frac{\kappa_{\mathrm{e,NR}}\beta_{\mathrm{2,NR}}f_{\mathrm{r}}}{\gamma_\mathrm{NR}}.
    \label{eq:Pcen}
\end{equation}

2.~3-dB bandwidth ($\Delta f_{\mathrm{3dB}}$):
\begin{equation}
    \Delta f_{\mathrm{3dB}} = 1.763\times\frac{\Delta\omega}{2\pi} =\frac{1.763}{\pi^2}\times \sqrt{-\frac{2n_0\delta\omega_{\mathrm{NR}}}{c\beta_{\mathrm{2,NR}}}}.
    \label{eq:3db_bandwidth}
\end{equation}
Here, the factor $1.763$ corresponds to $2\cosh^{-1}(\sqrt{2})$, which converts the spectral width from $1/\mathrm{e}$ to the full width at half maximum for a $\mathrm{sech}^2$-shaped spectrum. Combining Eqs.~\ref{eq:Pmin-detuning},~\ref{eq:3db_bandwidth}, we obtain a lower bound on the pump power necessary to sustain a soliton microcomb with a specified bandwidth:
\begin{equation}
   P_{\mathrm{in}} \geq -\frac{\pi^2}{1.763^2}\times\frac{\kappa_{\mathrm{NR}}\beta_{\mathrm{2, NR}}}{\eta_{\mathrm{NR}}\gamma_{\mathrm{NR}}}\times\frac{\Delta f_{\mathrm{3dB}}^2}{f_\mathrm{r}}, 
    \label{eq:Pmin}
\end{equation}
where $\eta_{\mathrm{NR}} = \kappa_{\mathrm{e,NR}} / \kappa_{\mathrm{NR}}$ is the loading factor. Equation~\ref{eq:Pmin} implies that the pump power requirement increases quadratically with spectral bandwidth and decreases with repetition rate. Dividing Eq.~\ref{eq:Pmin} by Eq.~\ref{eq:Pcen} removes material-specific parameters ($\beta_{\mathrm{2,NR}}, \gamma_\mathrm{NR}$) and yields an intrinsic constraint—termed the “impossible trinity”—that links the 3-dB bandwidth, repetition rate, and central-tooth power under available pump power, 
\begin{equation}
     \frac{P_{\mathrm{c}}\Delta f_{\mathrm{3dB}}^2}{f_{\mathrm{r}}^2} \leq 1.763^2\times \eta_{\mathrm{NR}}^2P_ {\mathrm{in}} \approx  3.1\times \eta_{\mathrm{NR}}^2P_ {\mathrm{in}}.
  \label{Eq:Trinity}
\end{equation}

\section{Theory for resonantly-coupled microresonators}

\subsection{Theoretical model and master equations}
The system comprising two coupled microresonators is described by a set of coupled LLEs:
\begin{equation}
    \frac{\partial B}{\partial T} = -\frac{\kappa_{\mathrm{RC}}}{2}B - i\delta \omega_{\mathrm{RC}} B +  \mathcal{F}\Bigl[iD_{\mathrm{int,RC}}(\mu)\tilde{B}_{\mu}\Bigr] +ig_{\mathrm{RC}}|A|^2A +iGA + \sqrt{\frac{\kappa_\mathrm{e,RC} P_\mathrm{in}}{\hbar\omega_0}},
\end{equation}
\begin{equation}
       \frac{\partial A}{\partial T}  = -\frac{\kappa_{\mathrm{NR}}}{2}A - i\delta \omega_{\mathrm{NR}} A +  \mathcal{F}\Bigl[iD_{\mathrm{int,NR}}(\mu)\tilde{A}_{\mu}\Bigr]+ig_{\mathrm{NR}}|A|^2A + iGB,
       \label{eq:lle_NR}
\end{equation}
where $B(T, \phi)$and $A(T, \phi)$ denote the slowly varying field amplitude of the resonant coupler (RC) and the nonlinear resonator (NR), respectively. The two quantities are normalized such that $|B|^2$ and $|A|^2$ correspond to the intracavity photon number. $\tilde{B}_{\mu}(T)$ and $\tilde{A}_{\mu}(T)$ are the optical field of the $\mu$-th mode, obtained from $B(T, \phi)$ and $A(T, \phi)$ via the Fourier transform, respectively. Definitions of these quantities: $\kappa_{\mathrm{RC(NR)}}, g_{\mathrm{RC(NR)}}$ are consistent with that in Section~\ref{sec:singletheory}. $D_{\mathrm{int,RC}}$ and $D_{\mathrm{int,NR}}$ are the integrated dispersion of the RC and NR, which are defined relative to the NR as $D_{\mathrm{int,RC(NR)}} = \omega_{\mu} - \omega_0 - \mu D_{1,\mathrm{NR}}$. The coupling strength between the microresonators is given by a real number $G$. $\delta\omega_{\mathrm{RC} (\mathrm{NR})}$ is the pump-RC (NR) detuning. $P_{\mathrm{in}}$ is the pump power on the RC. 

Notably, the NR and the RC in our configuration are evanescently coupled, with the coupling strength far below half of their average FSR \cite{yuan2023soliton}. In this regime, it is sufficient to consider coupling only between the 0-th modes of the two resonators, as other mode interactions have negligible impacts on the comb state in the NR. Furthermore, the Kerr nonlinearity in the RC can also be neglected as its intracavity energy remains well below the threshold for parametric oscillation. Thus, the RC is reduced to a single-mode linear microresonator. The resulting simplified model is given below when neglecting the Raman effect as well as third and higher-order dispersion terms of the NR:
\begin{equation}
    \frac{\partial b_{0}}{\partial T} = -\frac{\kappa_{\mathrm{RC}}}{2}b_{0} - i\delta \omega_{\mathrm{RC}} b_{0} +iGa_{0} + \sqrt{\frac{\kappa_\mathrm{e,RC} P_\mathrm{in}}{\hbar\omega_0}},
\end{equation}
\begin{equation}
       \frac{\partial A}{\partial T}  = -\frac{\kappa_{\mathrm{NR}}}{2}A - i\delta \omega_{\mathrm{NR}} A + i\frac{D_{\mathrm{2,NR}}}{2}\frac{\partial^2 A}{\partial \phi^2} +ig_{\mathrm{NR}}|A|^2A + iGb_{0},
       \label{eq:lle_NR}
\end{equation}
where $b_{0}$ and $a_{0}$ denote the field amplitude in 0-th mode of the RC and NR, respectively. 

For convenience, we normalize the coupled LLE as follows:
\begin{equation}
    \frac{\partial \psi_{\mathrm{RC}}}{\partial \tau}=-(\kappa_{\mathrm{r}} + i\zeta_{\mathrm{RC}})\psi_{\mathrm{RC}}+ig_{\mathrm{c}}\psi_{\mathrm{0,NR}}+f_{\mathrm{RC}},
\label{eq:lle_RC}
\end{equation}
\begin{equation}
    \frac{\partial \psi_{\mathrm{NR}}}{\partial \tau}=-(1 + i\zeta_{\mathrm{NR}})\psi_{\mathrm{NR}}+i d_{2,\mathrm{NR}}\frac{\partial^2 \psi_{\mathrm{NR}}}{\partial \phi^2}+i|\psi_{\mathrm{NR}}|^2\psi_{\mathrm{NR}}+ig_{\mathrm{c}}\psi_{\mathrm{RC}},
\label{eq:lle_NR}
\end{equation}
where $\tau = \frac{\kappa_{\mathrm{NR}}}{2}T$, $A = \sqrt{\frac{\kappa_{\mathrm{NR}}}{2g_{\mathrm{NR}}}}\psi_{\mathrm{NR}},b_{0} = \sqrt{\frac{\kappa_{\mathrm{NR}}}{2g_{\mathrm{NR}}}}\psi_{\mathrm{RC}},
a_0 = \sqrt{\frac{\kappa_{\mathrm{NR}}}{2g_{\mathrm{NR}}}}\psi_{\mathrm{0,NR}}, \zeta_{\mathrm{NR}} = \frac{2\delta \omega_{\mathrm{NR}} }{\kappa_{\mathrm{NR}}}, \zeta_{\mathrm{RC}}= \frac{2\delta \omega_{\mathrm{RC}} }{\kappa_{\mathrm{NR}}},\kappa_{\mathrm{r}} = \frac{\kappa_{\mathrm{RC}}}{\kappa_{\mathrm{NR}}},g_{\mathrm{c}} = \frac{2G}{\kappa_{\mathrm{NR}}},d_{2,\mathrm{NR}}= \frac{D_{\mathrm{2,NR}}}{\kappa_{\mathrm{NR}}},f_{\mathrm{RC}} = \sqrt{\frac{8g_{\mathrm{NR}}\kappa_{\mathrm{e,RC}}P_{\mathrm{in}}}{\kappa_{\mathrm{NR}}^3\hbar\omega_0}}$. The normalized pump power on RC can be described as
\begin{equation}
f_{\mathrm{RC}}^2=\frac{P_{\mathrm{in}}}{P_{\mathrm{th, NR}}}\cdot\frac{\kappa_{\mathrm{e,RC}}}{\kappa_{\mathrm{e,NR}}} = f_{\mathrm{NR}}^2\cdot\frac{\kappa_{\mathrm{e,RC}}}{\kappa_{\mathrm{e,NR}}}.
    \label{eq:fp}
\end{equation}
Here, $P_{\mathrm{th, NR}}$ is the parametric oscillation threshold of the NR when coupled from the bus waveguide, as defined in Section~\ref{sec:singletheory}, and $f_{\mathrm{NR}}^2$ denotes the corresponding normalized pump power for the NR.

\subsection{Effective pump power}
As shown in Eq.~\ref{eq:lle_NR}, the power injected into NR is predominantly governed by the coupling term $ig_{\mathrm{c}}\psi_\mathrm{RC}$. Therefore, we define the effective pump term as: 
\begin{equation}
f_\mathrm{eff}=ig_{\mathrm{c}}\psi_\mathrm{RC},
\end{equation}
such that Eq. \ref{eq:lle_NR} recovers the the standard form of the LLE. $|f_\mathrm{eff}^2|$ can be estimated by analyzing the steady-state continuous-wave solution of the coupled LLE. To this end, we first consider the optical field of the 0-th mode of the NR, given by,
\begin{equation}
    \psi_{\mathrm{0,NR}} \approx \frac{ig_{\mathrm{c}}\psi_{\mathrm{RC}}}{1+i\zeta_{\mathrm{NR}}}
\label{eq:cw_NR}.
\end{equation}
Inserting this into Eq. \ref{eq:lle_RC} and considering $\frac{\partial\psi_{\mathrm{RC}}}{\partial\tau}=0$, we can get:
\begin{equation}
    f_{\mathrm{RC}} = (\kappa_{\mathrm{r}}+i\zeta_{\mathrm{RC}})\psi_{\mathrm{RC}}+\frac{g_{\mathrm{c}}^2}{1+i\zeta_{\mathrm{NR}}}\psi_{\mathrm{RC}}.
    \label{eq:cw_RC1}
\end{equation}
Taking the modulus, we obtain the intracavity power of RC:
\begin{equation}
    |\psi_{\mathrm{RC}}|^2=\frac{f_{\mathrm{RC}}^2}{(\kappa_{\mathrm{r}}+\frac{g_{\mathrm{c}}^2}{1+\zeta_{\mathrm{NR}}^2})^2+(\zeta_{\mathrm{RC}}-\frac{g_{\mathrm{c}}^2\zeta_{\mathrm{NR}}}{1+\zeta_{\mathrm{NR}}^2})^2} .
\end{equation}
Thus, the effective pump power for the NR is given by
\begin{equation}
    |f_{\mathrm{eff}}^2| = g_{\mathrm{c}}^2|\psi_{\mathrm{RC}}|^2=\frac{g_{\mathrm{c}}^2f_{\mathrm{RC}}^2}{(\kappa_{\mathrm{r}}+\frac{g_{\mathrm{c}}^2}{1+\zeta_{\mathrm{NR}}^2})^2+(\zeta_{\mathrm{RC}}-\frac{g_{\mathrm{c}}^2\zeta_{\mathrm{NR}}}{1+\zeta_\mathrm{NR}^2})^2}.
    \label{eq:feff1}
\end{equation}
As the system transitions to the soliton state, large detuning in the NR causes, $\frac{g_{\mathrm{c}}^2}{1+\zeta_{\mathrm{NR}}^2} \to 0$, and the effective pump power converges to,
\begin{equation}
    |f_{\mathrm{eff}}^2| =\frac{g_{\mathrm{c}}^2f_{\mathrm{RC}}^2}{\kappa_{\mathrm{r}}^2+\zeta_{\mathrm{RC}}^2}.
\end{equation}
The effective pump power is maximized when the pump is resonant with the RC ($\zeta_{\mathrm{RC}} = 0$). In this regime, the optimized effective pump power is given by: 
\begin{equation}
    |f_{\mathrm{eff}}^2| =\frac{g_{\mathrm{c}}^2f_{\mathrm{RC}}^2}{\kappa_{\mathrm{r}}^2}.
    \label{eq:feff2}
\end{equation}

To quantify the advantage of employing the RC, we define an enhancement factor that compares the effective pump power delivered to the NR via the RC with the case using a conventional waveguide coupler, 
\begin{equation}
    \Gamma = \frac{|f_{\mathrm{eff}}^2|}{f_{\mathrm{NR}}^2}.
\end{equation}
According to Eq. \ref{eq:feff2} and Eq. \ref{eq:fp}, the enhancement factor can be calculated by
\begin{equation}
    \Gamma =\frac{4G^2}{\kappa_{\mathrm{NR}}\kappa_{\mathrm{RC}}}\cdot\frac{\eta_{\mathrm{RC}}}{\eta_{\mathrm{NR}}}.
\end{equation}
The loading factors $\eta_{\mathrm{NR} (\mathrm{RC})}=\kappa_\mathrm{e,NR(RC)}/\kappa_\mathrm{NR(RC)}$. For efficient coupling, they are usually in the range between 0.5 to 1. Therefore, the ratio $\frac{\eta_{\mathrm{RC}}}{\eta_{\mathrm{NR}}}$ is on the order of unity, and the enhancement factor is on the order of
\begin{equation}
    \Gamma\approx\frac{4G^2}{\kappa_{\mathrm{NR}}\kappa_{\mathrm{RC}}}.\label{eq:enhance}
\end{equation}

\subsection{Soliton existence range}
\label{sec:sol}
As discussed in previous works \cite{bao2014nonlinear,herr2014temporal}, the relationship between soliton detuning and pumping in a regular LLE is given by $\zeta \leq \frac{\pi^2 f^2}{8}$. Thus, for devices with RC, the maximum detuning range of the soliton is given by:
\begin{equation}
    \zeta_{\mathrm{NR}} \leq \frac{\pi^2 |f_{\mathrm{eff}}^2|}{8} = \Gamma\frac{\pi^2f_{\mathrm{NR}}^2}{8}.
    \label{Eq:zetaNRideal}
\end{equation}
Therefore, compared with waveguide-coupled NRs, in resonantly-coupled NRs the accessible detuning for soliton states is increased by a factor of $\Gamma$. However, as the detuning of the NR becomes large, a notable decrease of the effective pump power occurs. This requires additional correction that applies to the intracavity field of the NR (Eq. \ref{eq:cw_NR}), with contribution from the 0-th mode spectral component of the soliton. Adding it to the continuous-wave background gives:
\begin{equation}
    \psi_{\mathrm{0,NR}} = \sqrt{\frac{d_{\mathrm{2,NR}}}{2}}e^{i\varphi}+\frac{ig_{\mathrm{c}}\psi_{\mathrm{RC}}}{1+i\zeta_{\mathrm{NR}}}  \approx \sqrt{\frac{d_{\mathrm{2,NR}}}{2}}e^{i\varphi}+\frac{g_{\mathrm{c}}\psi_{\mathrm{RC}}}{\zeta_\mathrm{NR}}.
    \label{bgzero}
\end{equation}
where $\sqrt{\frac{d_{\mathrm{2,NR}}}{2}}e^{i\varphi}$ corresponds to the 0-th mode spectral component of the soliton and $\varphi$ corresponds to the phase of the soliton. At small detuning, the first term is negligible compared with the second term. However, at large detuning, its contribution should be considered. Substituting Eq. \ref{bgzero} into Eq. \ref{eq:lle_RC} and considering $\frac{\partial\psi_{\mathrm{RC}}}{\partial\tau}=0$ leads to:
\begin{equation}
    f_{\mathrm{RC}} +ig_{\mathrm{c}}\sqrt{\frac{d_{\mathrm{2,NR}}}{2}}e^{i\varphi}= (\kappa_{\mathrm{r}}+i\zeta_{\mathrm{RC}})\psi_{\mathrm{RC}}+\frac{g_{\mathrm{c}}^2}{1+i\zeta_{\mathrm{NR}}}\psi_{\mathrm{RC}}.
\end{equation}
Taking the modulus of both sides gives:
\begin{equation}
    |\psi_{\mathrm{RC}}|^2 =\frac{(f_{\mathrm{RC}}-g_{\mathrm{c}}\sqrt{\frac{d_{\mathrm{2,NR}}}{2}}\sin\varphi)^2+g_{\mathrm{c}}^2\frac{d_{\mathrm{2,NR}}}{2}\cos^2\varphi}{(\kappa_{\mathrm{r}}+\frac{g_{\mathrm{c}}^2}{1+\zeta_{\mathrm{NR}}^2})^2+(\zeta_{\mathrm{RC}}-\frac{g_{\mathrm{c}}^2\zeta_{\mathrm{NR}}}{1+\zeta_{NR}^2})^2}.
\end{equation}
As the detuning of the NR ($\zeta_{\mathrm{NR}}$) increases toward the upper boundary of the soliton existence range, we find that $\sin\varphi$ approaches 1, while $\cos\varphi$ tends to 0 \cite{herr2014temporal}: 
\begin{equation}
    \sin\varphi = \sqrt{\frac{8\zeta_{\mathrm{NR}}}{\pi^2g_{\mathrm{c}}^2|\psi_{\mathrm{RC}}|^2}} \to 1,\cos\varphi \to0.
\end{equation}
Therefore, the effective pump power for large detuning is modified to:
\begin{equation}
    |f_{\mathrm{eff}}^2| = g_{\mathrm{c}}^2|\psi_{\mathrm{RC}}|^2 \approx  g_{\mathrm{c}}^2\frac{(f_{\mathrm{RC}}-g_{\mathrm{c}}\sqrt{\frac{d_{\mathrm{2,NR}}}{2}})^2}{\kappa_{\mathrm{r}}^2},
\end{equation}
which further gives the modified the soliton existence range:
\begin{equation}
    \zeta_{\mathrm{NR}} \leq \frac{\pi^2}{8}\frac{g_{\mathrm{c}}^2(f_{\mathrm{RC}}-g_{\mathrm{c}}\sqrt{\frac{d_{\mathrm{2,NR}}}{2}})^2}{\kappa_{\mathrm{r}}^2} = \Gamma\frac{\pi^2f_{\mathrm{NR}}^2}{8}(1-\frac{g_{\mathrm{c}}\sqrt{\frac{d_{\mathrm{2,NR}}}{2}}}{f_{\mathrm{RC}}})^2.
\end{equation}
When the coupling strength $g_{\mathrm{c}}$ is moderate, the ratio $\frac{g_{\mathrm{c}}\sqrt{d_{\mathrm{2,NR}}/2}}{f_{\mathrm{RC}}}$ remains on the order of 0.1, and the enhancement factor remains nearly unaffected. As the coupling strength $g_{\mathrm{c}}$ increases such that $g_{\mathrm{c}}\sqrt{d_{\mathrm{2,NR}}/2}$ approaches $f_{\mathrm{RC}}$, the enhancement factor for the maximum detuning enabled by the RC configuration begins to diminish due to perturbations from the 0-th mode of soliton. This effect can be interpreted as an effective nonlinear loss induced by the soliton. It will be seen that such a refined theory exhibits better agreement with numerical simulations, as illustrated in Section~\ref{sec:sim}.

\subsection{Effective pump power in the modulation instability regime}
\label{sec:feffinmi}
The derivation above, which employs the coupled LLEs under steady CW conditions, captures the key physics of our system. However, this description assumes that the NR remains in either the CW or soliton state, where the intracavity field is well approximated by the steady-state solution. In MI regime, this assumption breaks down, as no well-defined steady-state solution exists for the intracavity field, making it difficult to directly determine $f_{\mathrm{eff}} $ using the previous approach. Nevertheless, once exceeding the parametric oscillation threshold, the zero-mode energy in the NR ($|\psi_{\mathrm{0,NR}}|^2$) fluctuates around 1, which is validated by the following simulations. This allows us to approximate $f_{\mathrm{eff}}$ as follows,
\begin{equation}
    f_{\mathrm{eff}}^2 \approx 1+(\zeta_{\mathrm{NR}}-1)^2
\end{equation}
Coincidentally, this result also corresponds to the boundary that separates the MI and soliton regions in the single-cavity phase diagram  \cite{godey2014stability,parra2014dynamics}.

\subsection{Phase diagram}
\label{sec:sim}
\begin{figure*}[h]
    \centering
    \includegraphics{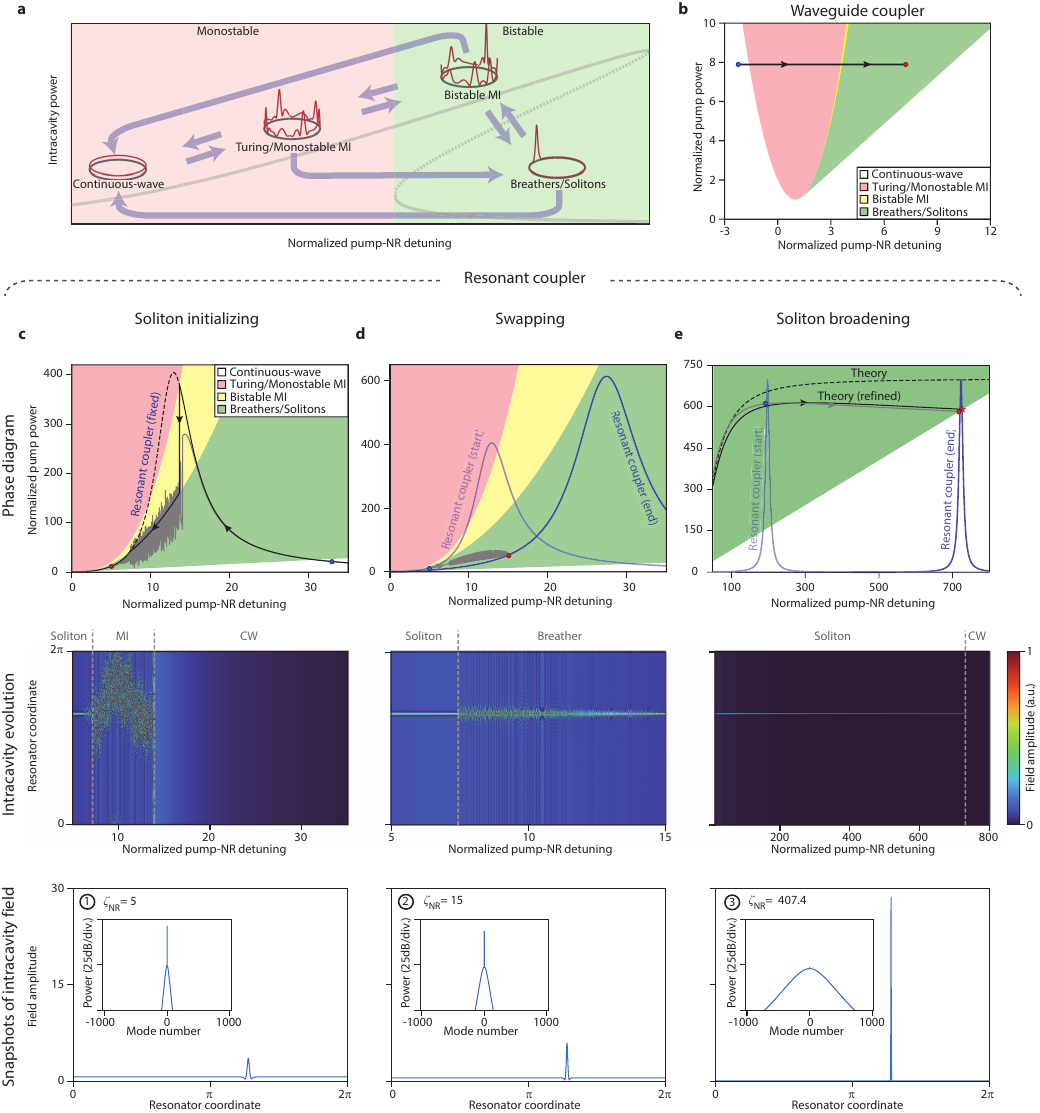}
    \caption{{\bf Theoretical and simulated pathways to soliton formation in phase diagrams.} {\bf a,} Schematic of the permissible transitions among distinct optical states in a nonlinear resonator (NR). Background shadings indicate the monostable and bistable regimes of the equilibrium state of the NR, suggesting that modulation instability (MI) can be classified accordingly. The presence of bistability influences how the system accesses the MI state~\cite{nielsen2021nonlinear} and soliton state. {\bf b,} Phase diagram for a waveguide-coupled NR, where the black trajectory delineates the evolution to soliton states at constant pump power. Blue and red dots mark the initiation and termination of the tuning process, respectively. {\bf c}--{\bf d,} Sequential stages for generating ultra-broadband solitons in a resonantly-coupled NR. Top panels: corresponding trajectories in the NR phase diagram. The blue curve represents the effective pump power delivered through the RC. The black and grey trajectories indicate the theoretical prediction and simulated evolution of the effective pump power during tuning, respectively. The simulated tuning trace begins and ends at the blue and red dots. The black dashed and solid curves in {\bf c} represent the effective pump power during tuning, predicted by the theory and the refined theory, respectively. The maximum detuning predicted by the refined theory is marked by the red star. Middle panels: temporal evolution of the intracavity field. Bottom panels: snapshots of the intracavity field at different pump-NR detunings. Insets: optical spectra.}
    \label{SIM1}
\end{figure*}

Here, we present the protocols to generate ultra-broadband soliton microcombs. The NR supports multiple optical states, yet direct access to the soliton state from the CW regime is unattainable \cite{herr2014temporal} (fig.~\ref{SIM1}a). Instead, soliton formation is initiated via modulation instability (MI), which in waveguide-coupled NRs is reached by precisely tuning the pump from the blue-detuned to the red-detuned side of the resonance (fig.~\ref{SIM1}b). This MI-to-soliton transition typically involves a sharp intracavity power drop and a concomitant thermo-optic blue-shift that can destabilize the soliton state; various techniques, including rapid power modulation, have been developed to address this issue \cite{yi2015soliton,brasch2016photonic}.

In resonantly-coupled NRs, the tuning mechanism is fundamentally altered. Soliton initiation begins by setting the RC frequency to be blue-detuned relative to the NR, while the pump, initially red-detuned from the NR, is tuned closer until MI is triggered (fig.~\ref{SIM1}c). The tuning process is conveniently visualized using a phase diagram defined by the relative detuning between the pump and the NR and the effective pump power delivered to the NR. In this diagram, the effective pump curve, determined by the fixed frequencies of the RC and NR during the initial tuning stage, initially leads to an increase in intracavity power as the pump is tuned closer to the NR resonance. The increased power drives the system into the monostable MI region, the only accessible pathway for transitioning from the CW state to MI~\cite{nielsen2021nonlinear}. As tuning continues, the effective pump power decreases, enforcing the transition into the soliton state. This tuning trajectory is effectively ``backward'' relative to the conventional approach. Notably, after soliton initiation using backward tuning, the pump resides on the thermally stable blue side of the hybrid resonance, which is favorable for subsequent tuning processes~\cite{Carmon:04}. To extend the comb span, the detuning is subsequently increased. This requires higher effective pump power while avoiding reentry into the MI regime. We achieve this by incrementally raising the NR's resonant frequency while lowering that of the RC until their frequencies effectively swap (fig.~\ref{SIM1}d). This maneuver shifts the effective pump curve, repositioning the soliton state to regimes of higher pump power and larger detuning. Following the swapping stage, we increase the NR's resonant frequency, while the pump laser remains aligned close to the RC resonance (fig.~\ref{SIM1}e). The increased detuning between the pump and the NR significantly broadens the soliton microcomb spectrum to an extent typically unattainable in waveguide-coupled devices. 

To validate our theoretical analysis, numerical simulations of the coupled LLE are performed using the Split-Step Fourier Transform method. Each optical mode is initially seeded with half the energy of a single photon. The normalized parameters used in fig.~\ref{SIM1} are: \(\kappa_{\mathrm{r}} = 6.52\), \(g_{\mathrm{c}} = 18.35\), \(d_{\mathrm{2,NR}} = 0.002\), and \(f_{\mathrm{RC}} = 9.277\).

We present the simulated evolution of states within the phase diagram, which delineates the emergence of nonlinear behavior in the NR. In fig.~\ref{SIM1}c, the black trajectory is the theoretical trace derived from the above analytical model, while the gray trajectory depicts the simulated RC intracavity energy scaled by \(g_{\mathrm{c}}^2\), corresponding to the effective pump power \(f_{\mathrm{eff}}^2\). The two curves agree well before the system approaches the boundary of the monostable MI regime, where the intracavity energy of the NR’s 0-th mode is far beyond the threshold for parametric oscillation. Continued tuning drives the system into the MI state, where the NR's 0-th mode energy abruptly drops and begins to fluctuate near 1. This sudden reduction causes the effective pump power to deviate from its original trajectory, exhibiting a sharp decline followed by oscillations near the MI–soliton boundary, as shown by the gray curve in fig.~\ref{SIM1}c and discussed in Section~\ref{sec:feffinmi}. The MI state persists until the effective pump power falls below a critical threshold, at which point a single soliton is generated, as illustrated in fig.~\ref{SIM1}c.

Once the soliton is generated, the swap operation begins by gradually tuning the RC resonance toward the pump laser, as previously discussed. After the swap, the effective pump power may not have reached its maximum (red point at the gray curve in fig.~\ref{SIM1}d). However, in the subsequent soliton broadening stage, tuning the NR mode further into the far red-detuned regime can also help approach the maximum effective pump power according to Eq. \ref{eq:feff1}. This simplifies the tuning process in the soliton broadening stage and is visualized in fig.~\ref{SIM1}e. Marked by the red star, the maximum detuning predicted by the refined theory (Section~\ref{sec:sol}) agrees well with the simulated result (red dot).

It should be emphasized that the discussion above is based on a simplified model considering only the group-velocity dispersion of the NR. When additional effects such as Raman scattering and higher-order dispersion are taken into account, the maximum detuning falls well below the ideal case (Eq.~\ref{Eq:zetaNRideal}), and so does the maximum spectral bandwidth~\cite{Parra-Rivas:14, PhysRevResearch.3.043207}. In other words, the enhancement factor inferred from the experimentally observed spectral bandwidth remains substantially below the ideal value predicted by Eq.~\ref{eq:enhance}. A complete model that incorporates additional effects is presented in Section~\ref{sec:sim2}.

\clearpage

\clearpage
\subsection{Simulation based on complete model for coupled resonators}
\label{sec:sim2}
The theoretical analysis and simulations above assume that the RC behaves as a single-mode linear microresonator. In practice, the RC can exhibit appreciable nonlinearity and dispersion and may also engage in additional mode coupling with the NR. To address these effects, we now introduce a comprehensive model that incorporates the RC’s intrinsic dispersion and Kerr nonlinearity, as well as Raman scattering and higher-order dispersion in the NR:
\begin{equation}
    \frac{\partial A}{\partial T}  = -i\delta \omega_{\mathrm{NR}} A  - \mathcal{F}\Bigl[\left(\frac{\kappa_{\mathrm{NR}}(\mu)}{2}+i D_{\mathrm{int,NR}}(\mu)\right)\tilde{A}_{\mu}- i G(\mu) \tilde{B}_{\mu}\Bigr] + i g_{\mathrm{NR}}|A|^2A - i g_{\mathrm{NR}}\,\tau_{\mathrm{R}} A \frac{\partial |A|^2}{\partial t} ,
\end{equation}
\begin{equation}
    \frac{\partial B}{\partial T}  = -i\delta \omega_{\mathrm{RC}} B  - \mathcal{F}\Bigl[\left(\frac{\kappa_{\mathrm{RC}}(\mu)}{2}+i D_{\mathrm{int,RC}}(\mu)\right)\tilde{B}_{\mu}- i G(\mu) \tilde{A}_{\mu}\Bigr] + i g_{\mathrm{RC}}|B|^2B  + \sqrt{\frac{\kappa_{\mathrm{e,RC}} P_{\mathrm{in}}}{\hbar\omega_0}}.
\end{equation}
To simulate the optical spectra of the high-power ultra-broadband soliton microcomb presented in the main text, we use the following parameters: \(Q_{\mathrm{0,NR}} = 6.48\)$\times 10^6$, \(Q_{\mathrm{0,RC}} = 6.75\)$\times 10^6$, \(g_{\mathrm{NR}} = 1.37\,\mathrm{Hz}\), \(g_{\mathrm{RC}} = 1.25\,\mathrm{Hz}\), \(\tau_{\mathrm{R}} = 0.45\,\mathrm{fs}\) and \(P_{\mathrm{in}} = 290\,\mathrm{mW}\). For the pump mode, \(Q_{\mathrm{e,NR}} = 3.81\)$\times 10^6$, \(Q_{\mathrm{e,RC}} = 0.38\)$\times 10^6$, \(G (0)/2\pi=1.65\,\mathrm{GHz}\). The coupling strengths-- both between the waveguide and the resonators, and between the resonators themselves-- are determined as a function of wavelength using finite-element simulations.

\begin{figure*}[h]
    \centering
    \includegraphics{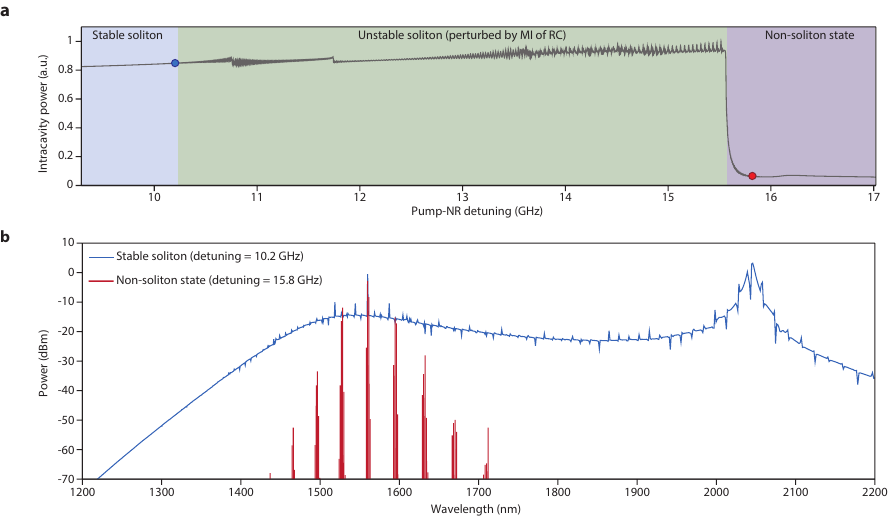}
    \caption{{\bf Simulation of high-power ultra-broadband soliton microcombs based on the complete model.} {\bf a,} Evolution of NR's intracavity power as the pump-NR detuning is increased. {\bf b,} Simulated optical spectra of the soliton state and non-soliton state indicated in \textbf{a}.}
    \label{SIM2}
\end{figure*}

The simulated maximum pump–NR detuning is approximately 10.2 GHz. Increasing the detuning beyond this limit induces modulation instability in the RC (fig.~\ref{SIM2}), which destabilizes the soliton microcombs in the NR.

\section{Additional experimental results}

\subsection{Resonant frequencies of coupled microresonators}
The hybridization of the pump resonances in the NR and RC is described by the following coupled equations: 
\begin{equation}
    \frac{\partial b_{0}}{\partial T} = -\frac{\kappa_{\mathrm{RC}}}{2}b_{0} - i\delta \omega_{\mathrm{RC}} b_{0} +iGa_{0} + \sqrt{\kappa_\mathrm{e,RC}}s_{\mathrm{in}},
\end{equation}
\begin{equation}
       \frac{\partial a_0}{\partial T}  = -\frac{\kappa_{\mathrm{NR}}}{2}a_0 - i\delta \omega_{\mathrm{NR}} a_0  + iGb_{0},
\end{equation}
Here, $s_{\mathrm{in}}=\sqrt{P_{\mathrm{in}}/\hbar\omega_0}$, and $|s_{\mathrm{in}}|^2$ represents the photon flux of the pump.
At the steady state, the intracavity field is given by:
\begin{equation}
    b_0 = \frac{\sqrt{\kappa_\mathrm{e,RC}}}{i\left(\delta\omega_{\mathrm{RC}}-\frac{G^2\delta\omega_{\mathrm{NR}}}{\delta\omega_{\mathrm{NR}}^2+\kappa_\mathrm{NR}^2/4}\right) + \frac{\kappa_\mathrm{RC}}{2}+ \frac{G^2\kappa_\mathrm{NR}/2}{\delta\omega_{\mathrm{NR}}^2+\kappa_\mathrm{NR}^2/4} }\cdot s_{\mathrm{in}}
    ,
\end{equation}
\begin{equation}
    a_0 = \frac{iG\sqrt{\kappa_\mathrm{e,RC}}}{i\left(\frac{\delta\omega_{\mathrm{RC}}\kappa_{\mathrm{NR}} + \delta\omega_{\mathrm{NR}}\kappa_{\mathrm{RC}}}{2}\right)-\delta\omega_{\mathrm{RC}}\delta\omega_{\mathrm{NR}} + \frac{\kappa_{\mathrm{RC}}\kappa_{\mathrm{NR}}}{4}+G^2 } \cdot s_{\mathrm{in}}.
\end{equation}
According to the input-output formalism, 
\begin{equation}
    s_{\mathrm{out, RC}}=-s_{\mathrm{in}}+\sqrt{\kappa_\mathrm{e, RC}}b_0,
\end{equation}
\begin{equation}
    s_{\mathrm{out, NR}}=\sqrt{\kappa_\mathrm{e, NR}}a_0, 
\end{equation}
where $s_{\mathrm{out, RC(NR)}}$ represents the output field at the through (drop) port. Thus, the transmission spectra at the through and drop are expressed as
\begin{equation}
    \left|\frac{s_{\mathrm{out, RC}}}{s_{\mathrm{in}}}\right|^2 = \frac{\left(\delta\omega_{\mathrm{RC}}-\frac{G^2\delta\omega_{\mathrm{NR}}}{\delta\omega_{\mathrm{NR}}^2+\kappa_\mathrm{NR}^2/4}\right)^2 + \left(\frac{\kappa_\mathrm{0, RC}-{\kappa_\mathrm{e,RC}}}{2}+ \frac{G^2\kappa_\mathrm{NR}/2}{\delta\omega_{\mathrm{NR}}^2+\kappa_\mathrm{NR}^2/4}\right)^2}{\left(\delta\omega_{\mathrm{RC}}-\frac{G^2\delta\omega_{\mathrm{NR}}}{\delta\omega_{\mathrm{NR}}^2+\kappa_\mathrm{NR}^2/4}\right)^2 + \left(\frac{\kappa_\mathrm{RC}}{2}+ \frac{G^2\kappa_\mathrm{NR}/2}{\delta\omega_{\mathrm{NR}}^2+\kappa_\mathrm{NR}^2/4} \right)^2},
\end{equation}
\begin{equation}
    \left|\frac{s_{\mathrm{out, NR}}}{s_{\mathrm{in}}}\right|^2 = \frac{G^2\kappa_\mathrm{e,RC}\kappa_\mathrm{e,NR}}{\left(\frac{\delta\omega_{\mathrm{RC}}\kappa_{\mathrm{NR}} + \delta\omega_{\mathrm{NR}}\kappa_{\mathrm{RC}}}{2}\right)^2+\left(\delta\omega_{\mathrm{RC}}\delta\omega_{\mathrm{NR}} - \frac{\kappa_{\mathrm{RC}}\kappa_{\mathrm{NR}}}{4}-G^2\right)^2 }.
\end{equation}
Based on the above equations, the calculated resonant frequencies of RC and NR relative to the pump for the three stages presented in Fig. 2d-f and Extended Fig Data. 1c-e are: $\delta\omega_{\mathrm{NR}}/2\pi=$ 0.34 GHz, $\delta\omega_{\mathrm{RC}}/2\pi=$ 3.27 GHz for soliton initializing, $\delta\omega_{\mathrm{NR}}/2\pi=$ 1.24 GHz, $\delta\omega_{\mathrm{RC}}/2\pi=$ 1.24 GHz for swapping, and $\delta\omega_{\mathrm{NR}}/2\pi=$ 8.9 GHz, $\delta\omega_{\mathrm{RC}}/2\pi=$ 0.3 GHz for soliton broadening. 

\subsection{Optical spectrum}

\begin{figure*}[h]
\centering \includegraphics{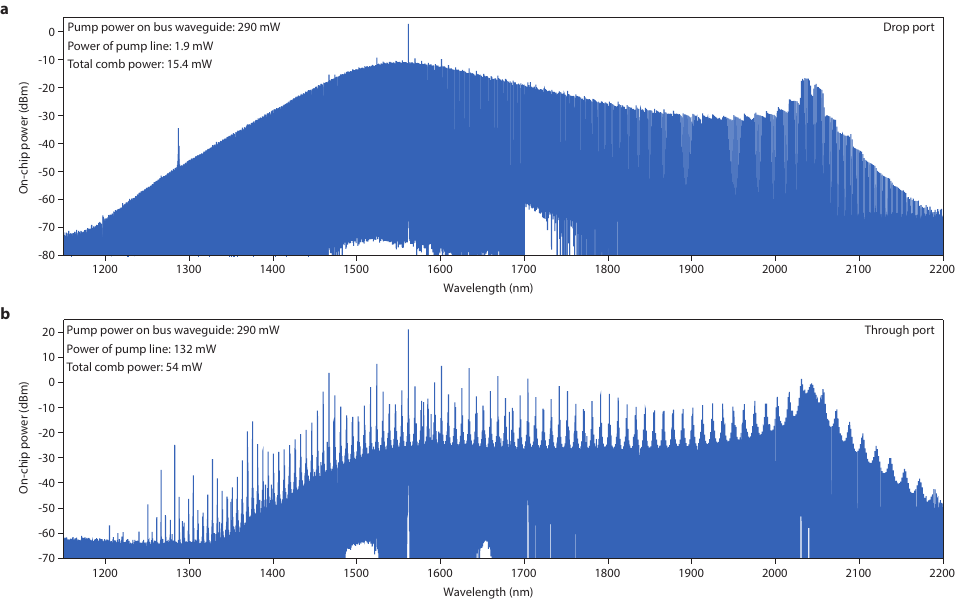} \caption{{\bf Optical spectra of the high-power ultra-broadband soliton microcomb.} \textbf{a,} Spectrum from the drop port. \textbf{b,} Spectrum from the through port.} \label{Combspec} \end{figure*}

Figure~\ref{Combspec} presents the optical spectra measured at both the drop and through ports while the NR operates in the high-power ultra-broadband soliton state, under a pump power of 290 mW applied to the bus waveguide. Notably, the spectrum obtained from the through port exhibits high-power teeth, which is attributed to mode crossings induced by the vernier effect between the resonances of NR and RC \cite{hu2022high}.

\subsection{Autocorrelation}

\begin{figure*}[h]
\centering \includegraphics{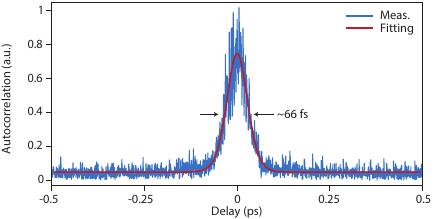} \caption{{\bf Intensity autocorrelation of the high-power ultra-broadband soliton microcomb.} The full-width-at-half-maximum of the autocorrelation trace is 66 fs.} 
\label{Autocorr}
\end{figure*}

The temporal profile of the high-power ultra-broadband soliton microcomb is characterized using an autocorrelator (APE pulseCheck), where the dispersion is compensated using a dispersion-compensating fiber. Fitting of the autocorrelation trace reveals that the soliton pulse exhibits a full-width-at-half-maximum (FWHM) of 42.7 fs (fig.~\ref{Autocorr}). It should be noted that the measured FWHM exceeds the 15.84 fs FWHM inferred from the optical spectrum (fig.~\ref{Combspec}a), likely due to the limited bandwidth offered by the frequency-doubling crystal in the autocorrelator.

\subsection{Measurement of the repetition rate}

\begin{figure*}[h]
\centering 
\includegraphics{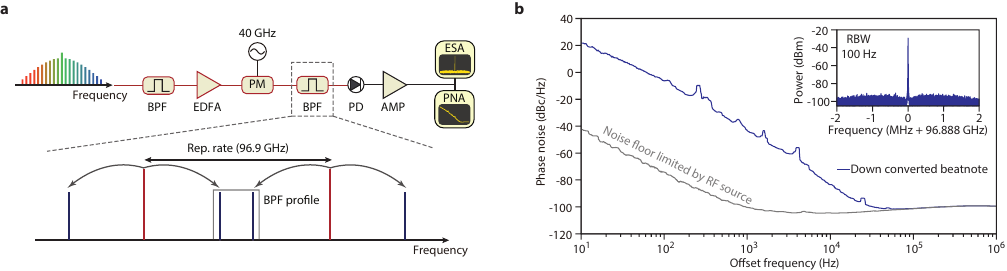} \caption{{\bf Coherence of the high-power ultra-broadband soliton microcomb.} \textbf{a,} Experimental setup for electro-optic (EO) downconversion. BPF: band-pass filter; EDFA: erbium-doped fiber amplifier; PM: phase modulator; PD: photodetector; AMP: electrical amplifier; ESA: electrical spectral analyzer; PNA: phase noise analyzer. The lower schematic illustrates the phase-modulated microcomb spectra entering the BPF, where EO-modulated sidebands (blue) of the adjacent microcomb modes (red) overlap to produce the downconverted beat note. \textbf{b,} Single-sideband phase noise of the downconverted beat note, along with the RF source contribution. Inset: beat note corresponding to the microcomb repetition rate. RBW: resolution bandwidth.} \label{Coherence} \end{figure*}

The repetition rate of the high-power ultra-broadband soliton microcomb is determined via electro-optic (EO) downconversion \cite{DelHaye2012Hybrid} (fig.~\ref{Coherence}a). In this measurement, two adjacent comb lines are isolated using a band-pass filter and subsequently modulated with a phase modulator operating at $f_{\mathrm{RF}} = 40$ GHz. This modulation generates sidebands around each comb line, and a pair of sidebands is selected by an additional band-pass filter. The resulting low-frequency beat note, $f_{\mathrm{beat}}$, is detected by a high-speed photodetector. By measuring the beat note using an electrical spectral analyzer, we can deduce the comb spacing, which is given by
\begin{equation}
f_{\mathrm{rep}} = 2f_{\mathrm{RF}} + f_{\mathrm{beat}}.
\end{equation}
In our experiments, $f_{\mathrm{beat}} = 16.888$ GHz, indicating a soliton microcomb repetition rate of $f_{\mathrm{rep}} = 96.888$ GHz. The phase noise of the downconverted beat note, $S_{\phi,\mathrm{beat}}$, is related to the phase noise of the repetition rate, $S_{\phi,\mathrm{rep}}$, and that of the RF source, $S_{\phi,\mathrm{RF}}$, according to \begin{equation} S_{\phi,\mathrm{beat}} = S_{\phi,\mathrm{rep}} + 4S_{\phi,\mathrm{RF}}. \end{equation} 
Figure \ref{Coherence}b presents the measured phase noise of the downconverted beatnote using a phase noise analyzer. For offset frequencies below 10 kHz, $S_{\phi,\mathrm{beat}}$ predominantly reflects $S_{\phi,\mathrm{rep}}$, as $S_{\phi,\mathrm{RF}}$ is comparatively negligible. At offset frequencies above 30 kHz, the noise is primarily limited by the RF source. These findings demonstrate the mutual coherence of the generated ultra-broadband soliton microcomb.

\subsection{Coherence of the dispersive wave}
\begin{figure*}[h] 
\centering 
\includegraphics{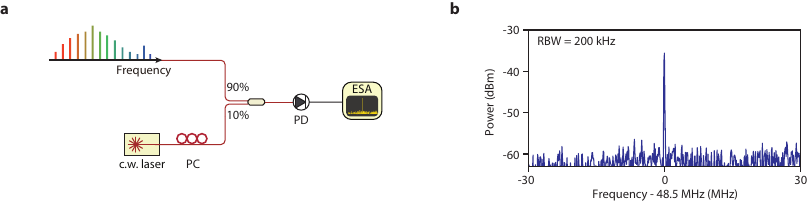} \caption{{\bf Heterodyne beat note measurements of the dispersive wave.} \textbf{a,} Experimental setup. PC: polarization controller. \textbf{b,} The measured beat note of the 100 GHz repetition-rate octave-spanning soliton microcomb with a narrow linewidth laser positioned at 1011.2 nm.} \label{DW} 
\end{figure*}

To assess the coherence of the dispersive wave of the 100 GHz repetition-rate octave-spanning soliton microcomb, we perform heterodyne beat note measurements. The soliton microcomb is combined with a tunable CW laser (Toptica CTL series) using a 90:10 fiber coupler. The combined signal is then detected by a silicon photodetector (New Focus Model 1801), which is responsive only to wavelengths below 1050 nm. In this configuration, only the dispersive wave components fall within the detector’s bandwidth. The resulting beat notes exhibit a narrow linewidth, confirming the coherence of the dispersive wave (fig.~\ref{DW}).

\newpage
\smallskip
\smallskip

\smallskip
\bibliography{SIref.bib}